\documentclass[11pt,preprint]{aastex}

\shorttitle{Evolution Accretion Disk NGC 1097}
\shortauthors{Storchi-Bergmann et al.}

\begin{document}

\title{Evolution of the Nuclear Accretion Disk Emission in NGC~1097:
Getting Closer to the Black Hole}

\author{Thaisa Storchi Bergmann\altaffilmark{1,2} and Rodrigo Nemmen da Silva}
\affil{Instituto de Fisica, UFRGS, Campus do Vale, Porto Alegre, RS, Brazil}
\email{thaisa@if.ufrgs.br}

\author{Michael Eracleous\altaffilmark{1,3,4}} \affil{Department of
  Astronomy and Astrophysics, Pensilvania State University, 525 Davey
  Lab, University Park, PA 16802}

\author{Jules P. Halpern\altaffilmark{1}} \affil{Columbia Astrophysics
  Laboratory, Columbia University, 550 West 120th Street, New York,
  NY 10027}

\author{Andrew S. Wilson\altaffilmark{1}} \affil{Astronomy Department,
  University of Maryland, College Park, MD 20742}

\author{Alexei V. Filippenko\altaffilmark{5,6}} \affil{Department of
  Astronomy, University of California, Berkeley, CA 94720-3411}

\author{Maria Teresa Ruiz\altaffilmark{1}} \affil{Departamento de
  Astronomia, Universidad de Chile, Casilla 36-D, Santiago, Chile}

\author{R. Chris Smith} \affil{Cerro Tololo
  Inter-American Observatory, National Optical Astronomy
  Observatories, Casilla 603, La Serena, Chile}

\and

\author{Neil M. Nagar\altaffilmark{1}} \affil{Arcetri Observatory, Largo
  E. Fermi 5, Florence 50125, Italy}

\altaffiltext{1}{Visiting Astronomer, Cerro Tololo Inter-American
  Observatory, which is operated by AURA, Inc.\ under contract to the
  National Science Foundation.}

\altaffiltext{2}{Visiting Astronomer, La Silla, European Southern
  Observatory, Chile.}

\altaffiltext{3}{Visiting Astronomer, Kitt Peak National Observatory,
  which is operated by AURA, Inc.\ under contract to the National
  Science Foundation.}

\altaffiltext{4}{Visiting Astronomer, MDM Observatory.}

\altaffiltext{5}{Visiting Astronomer, W. M. Keck Observatory, operated
  by Caltech, The University of California, and NASA.}

\altaffiltext{6}{Visiting Astronomer, Lick Observatory.}

\begin{abstract}

We study the evolution of the broad, double-peaked H$\alpha$
emission-line profile of the LINER/Seyfert 1 nucleus of
NGC~1097, using 24 spectra obtained
over a time span of 11 yrs --- from 1991 Nov. through 2002 Oct.
While in the first 5 yrs the main variation was in the relative
intensity of the blue and red peaks, in the last years we have
also observed an increasing separation between
the two peaks, at the same time as the integrated flux in the broad
line has decreased. We propose a scenario in which the emission
originates in an asymmetric accretion disk around a supermassive black
hole, whose source of ionization is getting dimmer, causing the
region of maximum emission to come closer to the center (and thus to
regions of higher projected velocity). We use the observations
to constrain the evolution of the accretion
disk emission and to evaluate two models: the elliptical disk model
previously found to reproduce the observations
from 1991 to 1996 and a model of a circular disk with
a single spiral arm. In both models the peak emissivity of
the disk drifts inward with time, while the azimuthal
orientation of the elliptical disk or the spiral
pattern varies with time. In the case of the spiral arm model,
the whole set of data is consistent with a monotonic precession
of the spiral pattern, which has completed almost two revolutions
since 1991. Thus, we favor the spiral arm model, which, through
the precession period implies a black hole mass that is consistent
with the observed stellar velocity dispersion. In contrast, the
elliptical disk model requires a mass which is an order of
magnitude lower. Finally, we have found tentative evidence of
the emergence of an accretion disk wind, which we hope to
explore further with future observations.

\end{abstract}


\section{Introduction}

The barred spiral galaxy NGC 1097 was the first example of a LINER
(low-ionization nuclear emission-line region; Heckman 1980) to display
broad, double-peaked Balmer lines. The double-peaked H$\alpha$ line
(FWHM $\approx 7500~{\rm km~s}^{-1}$) was discovered by
Storchi-Bergmann, Baldwin, \& Wilson
(1993; hereafter SB93) after a careful subtraction of the starlight
from the nuclear spectrum. The broad line was not easily recognizable
in the original spectrum because its contrast against the continuum
from the underlying stellar population was fairly low.  This discovery
was followed by other examples, all of them in 
{\it Hubble Space Telescope (HST)} spectra taken
through narrow apertures in which starlight from the host galaxy is
largely excluded.  These examples include M81 (Bower et al. 1996),
NGC~4203 (Shields et al. 2000), NGC~4450 (Ho et al. 2000), and NGC~4579
(Barth et al. 2000).

Comparison of the double-peaked H$\alpha$ spectrum of NGC~1097 with
earlier observations, as well as the later evolution of the
double-peaked profile, led to the conclusion that the broad Balmer
lines were associated with a transient event (e.g., SB93, Eracleous et
al. 1995; Storchi-Bergmann et al. 1995, 1997; hereafter SB95, SB97),
presumably an episode of mass accretion onto the nuclear supermassive
black hole (SMBH).

The recent findings of a proportionality between the masses of
galactic bulges and that of the SMBH in galaxies that are close enough
to allow the SMBH's detection via stellar kinematics (e.g., Ferrarese \&
Merritt 2000; Gebhardt et al. 2000) supports their presence in the
nucleus of most present-day galaxies. Such a SMBH
will eventually accrete mass, such as via the capture of a star passing
closer to the black hole than its tidal radius.  
If the SMBH has a mass smaller than $\sim10^8
{\rm M}_{\odot}$, a solar-mass star will be tidally disrupted before
it is accreted, leading to the formation of an accretion disk or ring.
As argued by SB95 and SB97, such a process is the preferred
interpretation for the double-peaked lines of NGC~1097 and may also
be applicable to some of the other LINERs mentioned above.  It is thus possible
that we have begun to witness such tidal disruption events, which have
long been predicted by theory (see review by Rees 1988).

Monitoring the variability of the double-peaked profiles provides
useful constraints on the above scenario. Therefore, we have been
following the evolution of the H$\alpha$ profile of NGC~1097 since it
was first discovered in 1991. In this paper we present the result 
of 11 years of monitoring and use the profile observations to constrain and
refine scenarios for their origin, such as the precessing elliptical
ring suggested by Eracleous et al. (1995) and SB95.

In \S2 we describe the observations and analysis of the data, in \S3
we present the results, and in \S4 we discuss their interpretation and
present two possible accretion disk models to reproduce the profile
variations. We explore in \S5 the significance of our findings and
compare them with those for other objects. The
conclusions of this work are presented in \S6.

\section{Observations}

We observed the nuclear spectrum of NGC~1097 from 1991 to 2002, up to
several times a year, using mostly the CTIO 4m Blanco and 1.5m
telescopes, and the KPNO 2.1m telescope. We also obtained a few
spectra with the Keck II, Lick 3m, MDM 2.4m, and ESO NTT
telescopes. The log of observations is given in Table \ref{tbl-1},
where we also list the slit and extraction window widths.

The first observations (SB93, SB95, SB97) were obtained through a
2\arcsec\ slit and an extraction window of 4\arcsec. We later used a
narrower slit of 1\farcs5 for the observations with the large
telescopes (3.6m and 4m) in order to improve the contrast between the
double-peaked profile and the starlight spectrum, but keeping it wide
enough to avoid spurious variations due to seeing differences. We have
also kept the extraction window large enough to allow the
normalization of the double-peaked broad line flux using the fluxes of
the narrow emission lines, as discussed below.

In order to isolate the double-peaked profile, we subtracted the
contribution of the stellar population, as described in SB93, SB95, and
SB97. In the case of the observations with the Blanco and NTT
telescopes, the stellar population spectrum was extracted from windows
adjacent to the nuclear window, typically from regions between
3\arcsec\ and 6\arcsec\ from the nucleus. The nuclear spectrum is
redder than that of the stellar population at these windows.
Attributing the difference in color to reddening by dust, we estimate
a value of $E(B-V) \approx 0.05$ mag. We have thus reddened the stellar
population spectra accordingly and normalized them to the flux of the
continuum in the nuclear spectrum. In addition, we have excized the
residual narrow-line emission from the stellar population spectra
(using as a reference spectra of disk galaxies free of emission lines), 
and have subtracted it from the nuclear spectrum.  The
spectra obtained with the smaller telescopes (CTIO 1.5m and KPNO
2.1m) have a signal-to-noise ratio too low outside the nucleus to allow the
extraction of a population template; for these we used the 
starlight spectra extracted from the CTIO 4m observations to
subtract the stellar population contribution. The same was done for the Keck
and MDM spectra. The process of starlight subtraction is more
important in the most recent spectra because the double-peaked line is
fading (see discussion below) and its contrast to the stellar
population spectrum is decreasing.  Thus a careful match of the stellar
population spectrum is necessary before the subtraction.

\begin{figure}
\plotone{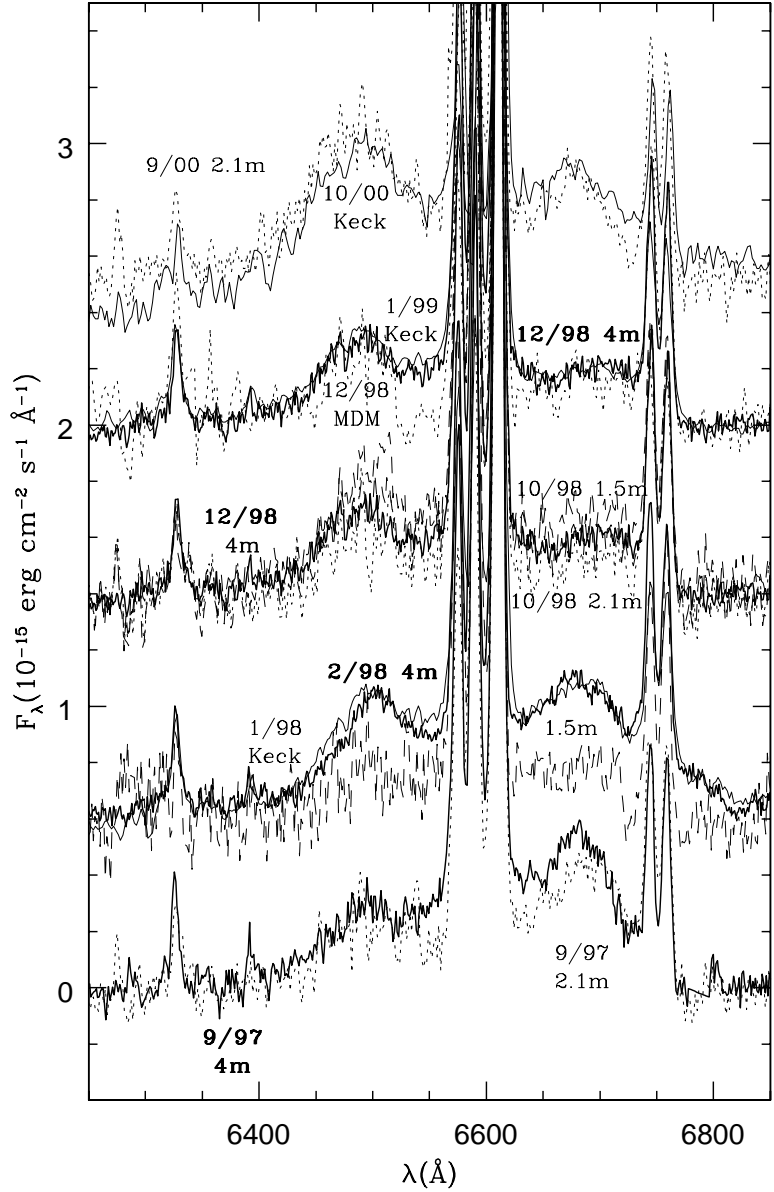}
\caption{Comparison between profiles obtained at approximately the
  same epoch (within two months) but using different
  telescopes. Spectra are identified in the figure by the epoch
  (mm/yy) and telescope: CTIO 4m (heavy continuous line), Keck II
  (continuous line), KPNO 2.1m (dotted line), MDM for epoch 12/98
  (dotted line as well), and CTIO 1.5m (dashed line).\label{fig1}}
\end{figure}

The last step was to normalize the broad-line flux, which we have done
using the fluxes of the narrow emission lines H$\alpha$,
[\ion{N}{2}] $\lambda\lambda$6548, 6584, and
[\ion{S}{2}] $\lambda\lambda$6717, 6731, under the assumption that
they have not varied during the 11 years of observations. This is a
reasonable assumption, considering that the average extraction windows
correspond to $\sim 150 \times 300$ pc$^2$ at the galaxy.
For the normalization we used as reference the spectrum of 1996 Jan. 24,
as follows. For each spectrum we determined the average ratio of the
[N~II] + H$\alpha$(narrow) and [S~II] fluxes with respect to the 1996 Jan. 24
spectrum and multiplied the spectrum by this average ratio.
This scaling factor is listed for each spectrum in the last
column of Table \ref{tbl-1}. The success of this
procedure is illustrated in Figure \ref{fig1}, which shows that profiles
obtained approximately at the same epoch (within two months of
each other) using different telescopes match each other
very well after the above normalization. This result
also shows that the small differences
in apertures of the different spectra does not significantly affect
the normalization.
 
The Keck spectra are the only exceptions, as we had to ``re-normalize''
them to $\sim$80\% of the resulting flux after the normalization
described above, to allow an agreement with the spectra obtained
at approximately the same epoch with the other large telescopes.
We attribute this result to the narrower slit used in the Keck spectra
($1''$), suggesting that the narrow-line emission extends
beyond 1\arcsec\ although not much more, as the normalization works for
the larger apertures. The scaling factors listed
in the last column of Table \ref{tbl-1} for the Keck spectra
are already corrected for this additional normalization.

After normalization, Figure \ref{fig1} shows that there is reasonable agreement
between the CTIO 4m and KPNO 2.1m spectra of 1997 Sep. There is
excellent agreement between the spectra with the best signal-to-noise
ratios: Keck (thin continuous line; epochs 1998 Jan. and
1999 Jan.) and CTIO 4m (thick continuous lines; epochs
1998 Feb. and 1998 Dec.). In comparison, the MDM spectrum of 1998 Dec.
shows a broad dip around 6530~\AA\ which is not present in the
Keck and CTIO spectra.  There is also good agreement between the Keck
spectra of 2000 Oct. and the KPNO 2.1m spectra of  2000 Sep.  The
spectra obtained with the CTIO 1.5m have the poorest signal-to-noise
ratios: the profile from 1998 Jan. has very little flux and is not
compatible with the profiles measured both with the Keck and CTIO 4m
at approximately the same epoch. On the other hand the 1.5m profile of
1998 Oct. agrees with that of the KPNO one of 1998 Sep. and with the
CTIO 4m one of 1998 Dec.

On the basis of Figure \ref{fig1} we conclude that we should give more
weight in our analysis to the results based on the Keck, CTIO 4m,
and NTT 3.6m spectra relative to the ones obtained with
the smaller telescopes. This figure also illustrates that
changes in the shape of the double-peaked
profiles seem to occur on time scales longer than a few ($\sim$
3--4) months.

\section{Results}

\begin{figure}
\plotone{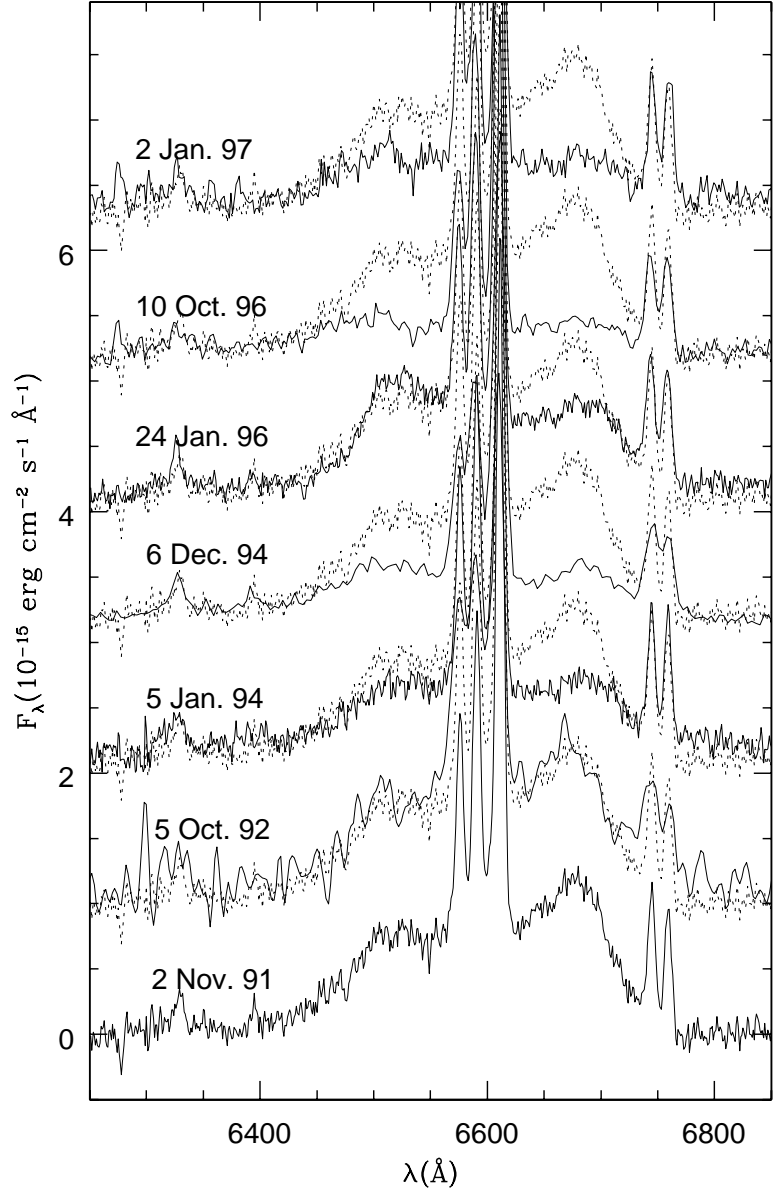}
\caption{Broad double-peaked H$\alpha$ profiles from 1991 Nov. to
  1997 Jan. (after subtraction of the stellar population
  contribution). Dotted lines show the 1991 Nov. profile
  superimposed on the subsequent profiles, for
  comparison.\label{fig2}}
\end{figure}

\begin{figure}
\plotone{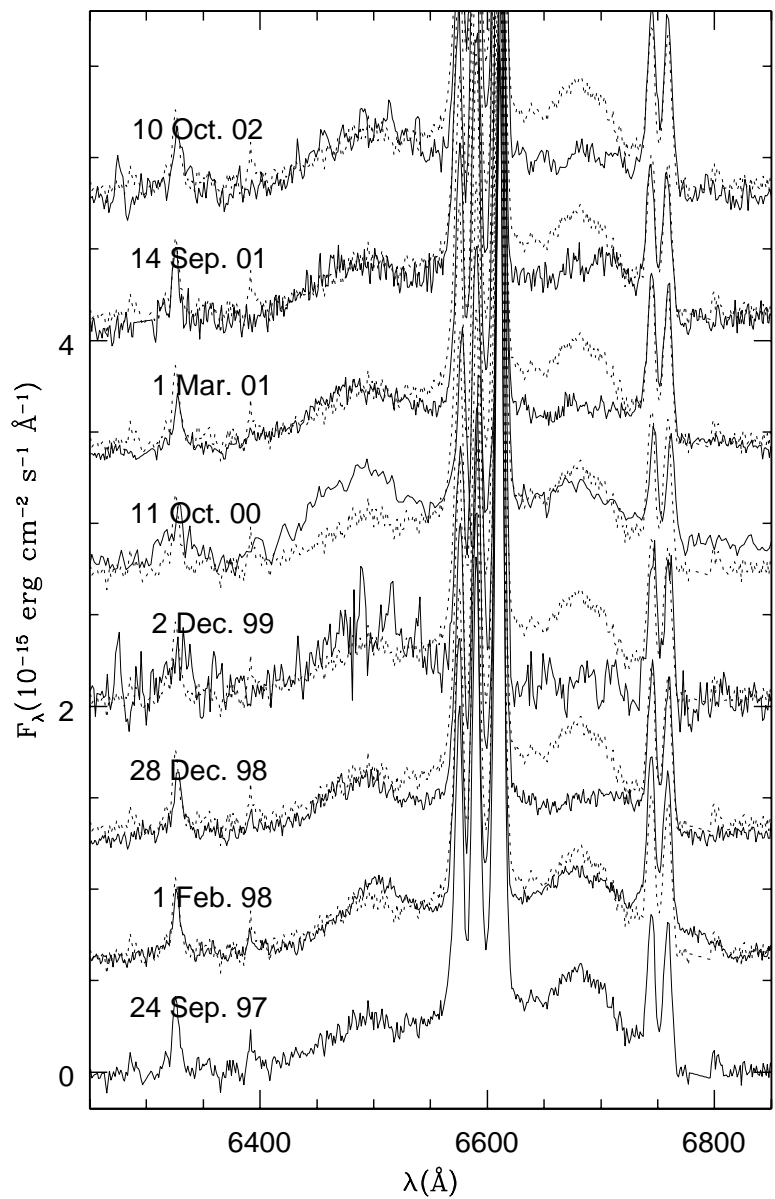}
\caption{Broad double-peaked H$\alpha$ profiles from 1997 Sep. to
  2002 Oct. (after subtraction of the stellar population
  contribution). Dotted lines show the 1997 Sep. spectrum superimposed
  on the subsequent profiles, for comparison.\label{fig3}}
\end{figure}

In Figures \ref{fig2} and \ref{fig3} we show a sequence of selected
H$\alpha$ profiles, from 1991 Nov. 2 to 2002 Oct. 10. Our
criteria for the selection was to obtain the most complete time
coverage using the available data, but using spectra having the 
highest signal-to-noise ratios in cases of multiple spectra at a
given epoch, as discussed above.

During the evolution of the double-peaked H$\alpha$ profile shown in
Figures  \ref{fig2} and \ref{fig3}, the most noticeable changes occur in
the relative strengths of the blue and red peaks. The profile starts
with the red peak stronger than the blue in 1991 Nov. and evolves to
similar strengths of the two peaks in 1994 Jan., then to the blue peak
stronger than the red in 1996 Jan. The spectrum of 1997 Sep. shows
again the red peak stronger than the blue, then the two peaks show
similar strengths by mid 1998, and then the blue peak is 
stronger than the red by 1999.  In the
observation of 2001 Sep. the blue and red peaks have similar
strengths, and in 2002 Oct. the blue peak is stronger again. Another
important change that could only be noticed after accumulating data
over so many years is a {\it decrease of the broad-line flux at the same
time as the blue and red peaks are moving progressively farther
apart}, as indicated by the measurements below.

In order to quantify the observed variations, we measured the
wavelengths of the blue and red peaks, their peak flux densities, and
the flux of the broad line, which we list in Table \ref{tbl-2}. The
wavelengths of the blue and red peaks ($\lambda_B$ and $\lambda_R$)
and corresponding peak flux densities ($F_B$ and $F_R$) were obtained
by fitting Gaussians separately to the blue and red peaks of the
profiles and measuring the central wavelengths and peak fluxes. These
measurements were only possible in the large-telescope spectra, as the
fits to the blue and red peaks did not consistently converge to a
unique central wavelength in spectra with low signal-to-noise ratio.
The flux of the broad line ($F_{broad}$) was measured by integrating
the flux of the broad plus the narrow emission lines and then
subtracting the fluxes of the narrow lines.  Uncertainties in the
wavelengths and fluxes were estimated by repeating the fits several
times after changing by a few angstroms the limits of the windows used
to fit the Gaussians and to measure the broad-line flux. The resulting
average uncertainties are $\sim 5$~\AA\ for $\lambda_B$ and
$\lambda_R$, $\sim 5\times 10^{-17}~{\rm
erg~cm^{-2}~s^{-1}~\AA^{-1}}$ for $F_B$ and $F_R$, and $\sim
2\times 10^{-14}~{\rm erg~cm^{-2}~s^{-1}}$ for $F_{broad}$.

Figure \ref{fig4} shows the temporal variations of the blue and
red peak velocities ($V_B$ and $V_R$), obtained from
$\lambda_B$, $\lambda_R$, and  the average wavelength of the narrow
H$\alpha$ line as reference (6591.7~\AA), as well as the temporal variation of
$V_R - V_B$ and of $F_{broad}$. We observe a blueshift of the
blue peak from $\sim -2800~{\rm km~s}^{-1}$ in the first years to
$\sim -4500~{\rm km~s}^{-1}$ in 2001, and a redshift of the
red peak from $\sim 3500~{\rm km~s}^{-1}$  to 
$\sim 4500~{\rm km~s}^{-1}$ over the same period. The resulting
$V_R - V_B$ increases from $\sim 6000~{\rm km~s}^{-1}$ to
$\sim 9000~{\rm km~s}^{-1}$. In Figure \ref{fig5} we show
the temporal variation of $F_B/F_R$.
 
\section{Interpretation and Comparison With Models}

\subsection{Preliminary Inferences from the Data\label{interp}}

The continued variation of $F_B/F_R$ and, in particular, the strong
red peak observed in 1997 Sep. still support the origin of the
profile in an accretion disk that is not axisymmetric.  The asymmetry
could be the result of an eccentricity (e.g., Eracleous et al. 1995;
SB95), or a circular disk with a disturbance, such as a bright spot
(Newman et al. 1997; Zheng, Veilleux, \& Grandi 1991) 
or spiral arm (e.g., Chakrabarti \& Wiita 1993;
Gilbert et al. 1999). The precession of the elliptical disk or of the
disturbance in the circular disk would be the origin of the variation
of $F_B/F_R$ with time.

The increasing separation of the blue and red peaks (\S3) indicates that the
bulk of the broad-line emission is coming from regions of successively
larger velocities.  In the accretion disk scenario, this means that
the bulk of emission is coming from successively smaller radii.  The
above result could suggest at first glance that we are witnessing the
infall of matter toward the SMBH. Unfortunately, this seems not to
be possible over ten years of observations. A noticeable decrease in
radius should occur in the viscous time scale, which should be larger
than 10$^4$ years in the case of NGC~1097 (Frank, King, \& Raine 1992,
p. 99, using the black hole mass obtained in Sec. \ref{s_mass}).

An alternative possibility is borne out by the observation that, at
the same time that the profile is getting broader, the line flux is
decreasing (in 2001 Sep. the broad H$\alpha$ flux is
$\sim$1/3 of the flux of 1991 Nov.), which suggests that
the ionizing source is getting dimmer. Thus, a more plausible
interpretation is that the dimming of the ionizing source is causing
the region of maximum emission to come closer to the center, to regions
of larger velocities and thus giving rise to broader profiles.

\begin{figure}
\plotone{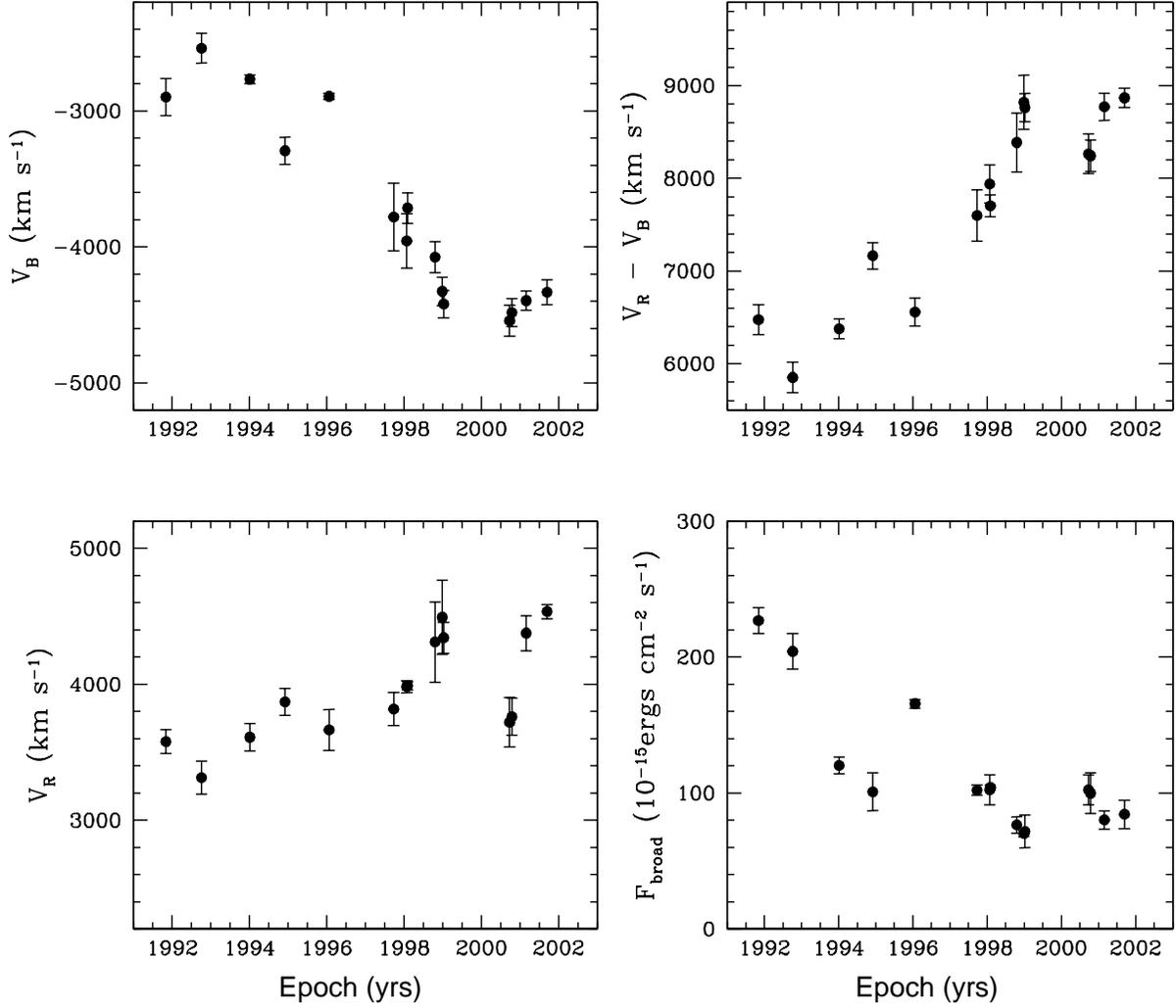}
\caption{Evolution of the measured properties of the double-peaked
  profiles. Left: Peak velocities corresponding to Gaussian curves fitted to the
  blue ($V_B$, top) and red ($V_R$, bottom) peaks. Right: $V_R - V_B$ (top) and
$F_{broad}$ (bottom).\label{fig4}}
\end{figure}

\begin{figure}
\plotone{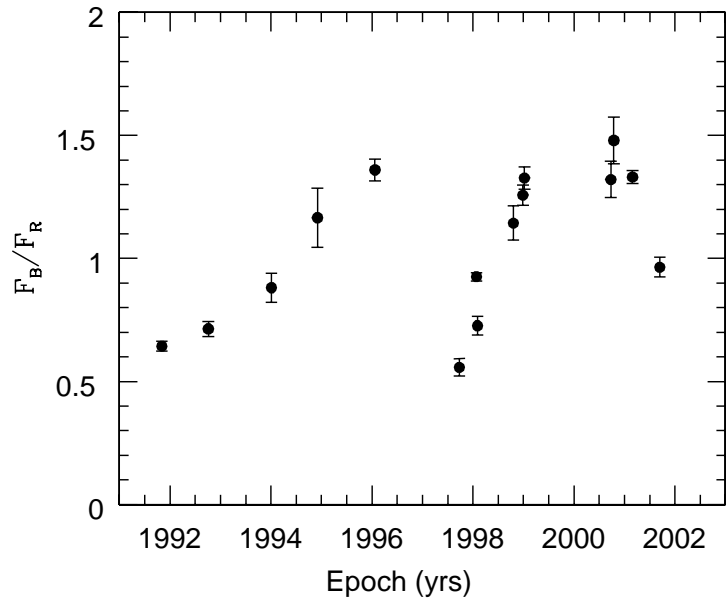}
\caption{Evolution of ratio between the
  blue and red peak fluxes (of the fitted Gaussians). \label{fig5}}
\end{figure}

\begin{figure}
\plotone{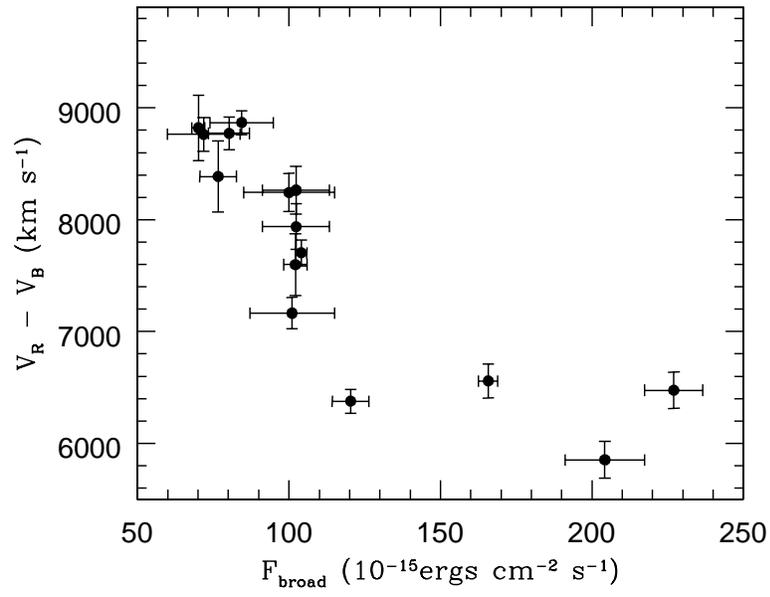}
\caption{The inverse correlation between the velocity difference between the two
peaks and the flux of the broad H$\alpha$ line. \label{fig6}}
\end{figure}

Inspection of Figure \ref{fig4} shows that there seems indeed
to be an inverse correlation between $V_R - V_B$ and $F_{broad}$,
not only as a general trend along the 11 years covered
by the observations, but also from epoch to epoch.
For example, in 1996 Jan. the flux of the line showed an
increase relative to the two previous epochs, while
$V_R - V_B$ decreased. The same seems to have happened in Sep.--Oct.
2000. This inverse correlation can be better
observed in Figure \ref{fig6}, where we plot $V_R - V_B$
against $F_{broad}$. 

The above effect (the inverse correlation between the
ionizing flux and width of the profile) has been
previously discussed by Dumont \& Collin-Souffrin (1990).  These
authors explain the existence of a ``radius for maximum emission''
or ``saturation radius'' as follows. In the line-emitting region of
the accretion disk there are two effects contributing to the line
intensity: the increase of the emitting surface with distance from the
center, and the decrease of the intensity of ionizing radiation with
radius. The ionizing flux can become quite high close to the center,
leading to ionization of the disk and saturation of the line emission
(cooling is mainly due to free-bound and free-free processes and not
to lines). The radius at which this saturation sets in is the radius
of maximum emission.

Finally, we notice that in Figure \ref{fig4},
the velocity of the blue peak shows a steeper variation (blueshift)
than that of the red peak (redshift), implying a net blueshift of the center
between the two peaks of $\sim -450~{\rm km~s}^{-1}$
between 1991 Nov. and 2001 Sep. One possibility is that
this blueshift is due to a wind originating in the disk,
which would be consistent with observations of Fe~II absorption
lines in a recent {\it HST} spectrum of NGC~1097 (Eracleous 2002)
which indicate the presence of an outflow.

With these considerations in mind, we have tried to fit the variable
line profiles with two different models, as we describe in detail
below. Both models rely on a non-axisymmetric disk structure and its
precession to reproduce the changes in the relative strengths of the
two peaks. However, this mechanism cannot explain the increasing
separation of the two peaks. The latter effect requires us to change
the radius of maximum line emission in the disk by adjusting the
prescription for the disk emissivity.

\subsection{Elliptical Disk Model\label{s_ell}}

In Eracleous et al. (1995) we showed that an elliptical accretion ring
model could reproduce the broad double-peaked H$\alpha$ profile of
NGC~1097, as observed in 1991 Nov.  After following the
variations of the line profile from 1991 to 1996 (SB97), we developed
the model further so that the plane of the ring was inclined at
$i=34^{\circ}$ to the line of sight, and its line-emitting part
corresponded to a range of pericenter distances between
$\xi_1 = 1300$ and $\xi_2 = 1600$, where $\xi$ is
expressed in units of the gravitational radius $r_g\equiv GM/c^2$. The
ring was circular up to a radius $\xi_c = 1400$, and from there the
eccentricity increased linearly from $e = 0$ to 0.45.  The ring had a
broadening parameter $\sigma=1200~{\rm km~s}^{-1}$ and an emissivity
varying as $\epsilon(\xi)=\epsilon_0\; \xi^{-q}$. The
variation of the observed line profiles from 1991 to 1996 were
reproduced by the precession of the ring with a period of of 25 years,
which led to an estimate of the central black hole mass of $9\times
10^5~{\rm M}_\odot$.

We tried to use the same elliptical ring model to reproduce all the
double-peaked H$\alpha$ profiles observed from 1991 to 2002, by
changing only the orientation angle, $\phi_0$, under the assumption
that the precession continued. However, we were not able to reproduce
the line profiles observed after 1996; continued precession alone
could explain the variation in the relative heights of the two peaks,
but not the observed velocity shifts of the blue and red peaks.  After
exploring the effect of other parameters on the shape of the line
profile, we found that only changes in the inner pericenter distance
of the ring (from $\xi_1=1300$ to 450) could lead to the
observed increase in the separation of the two peaks, $\lambda_R -
\lambda_B$. Such a dramatic change in the physical location of a given
volume of gas could not occur on a time scale of
5 years, however, as we noted in the previous section. Rather, the
emission properties of the ring must have changed so that the radius
of maximum emission became smaller.  Therefore, to produce a
self-consistent and physically plausible scenario for the entire set
of observations, we were led to revise the model parameters. This
revision represents a refinement of earlier models, which is possible
only after following the variability of the line profile over a
baseline of 10 years.  We now propose that the ring has always had a
smaller inner radius of $\xi_1=450$ (and now resembles a disk more
than a ring), and that the increasing width of the profile is due to a
change in emissivity such that, in more recent epochs, the inner
regions of the disk make a more significant relative contribution to
the total line emission. This picture is also consistent with the
observational result that the flux of the broad line is decreasing,
as discussed above.

To explain the increasing separation between the two peaks, we
experimented with several emissivity laws. We found that we could best
reproduce the observations with a broken power law, such that inside a
radius $\xi_q$ the power-law index is $q_1=-1$ and outside this radius
$q_2=1$. This means that the emissivity {\it increases} with radius
from $\xi_1$ to $\xi_q$ and then {\it decreases} with radius beyond
$\xi_q$. These $q$ values are also physically motivated: in the
``saturation'' region, the increase in emissivity is due to
the increase in the area of the ring, while beyond $\xi_q$, the
decrease can be understood as due to the radiation field dropping
off as $r^{-2}$ and the increase of the emitting area as $r^1$.

 In this parameterization of the disk emissivity, varying the
``break radius'' $\xi_q$ allows us to reproduce the shifts of the
blue and red peaks.  In particular, allowing the ``break radius'' to
decrease with time shifts the region of maximum emission closer to the
center and results in an increase in the separation between the two
peaks. At the same time, since the relative heights of the two peaks
have continued to change, we allow the disk to precess (as in SB97).

\begin{figure}
\plotone{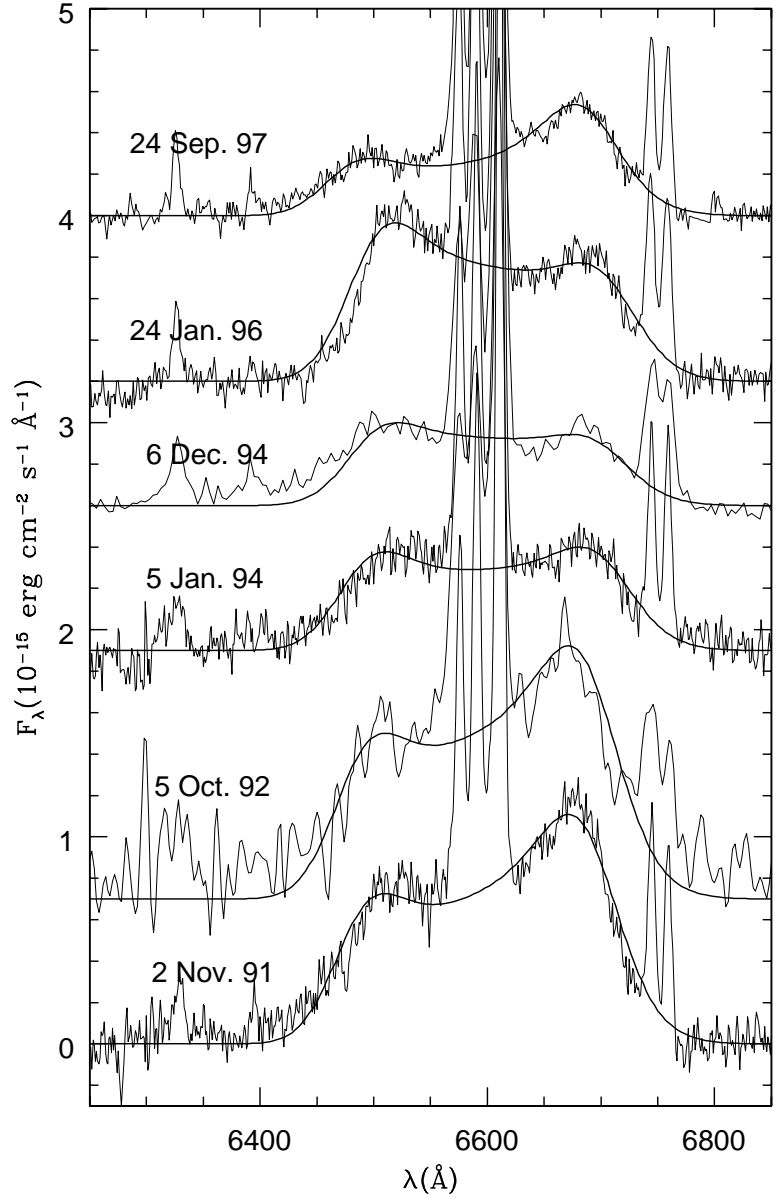}
\caption{Fits of the elliptical disk model to the profiles from
  1991 Nov. to 1997 Sep. by changing only $\phi_0$ (the orientation of
  the disk relative to the line of sight) and $\xi_q$ (the radius of
  the peak of the emissivity law). The values of these parameters are
  listed in Table \ref{tbl-3}. \label{fig7}}
\end{figure}

\begin{figure}
\plotone{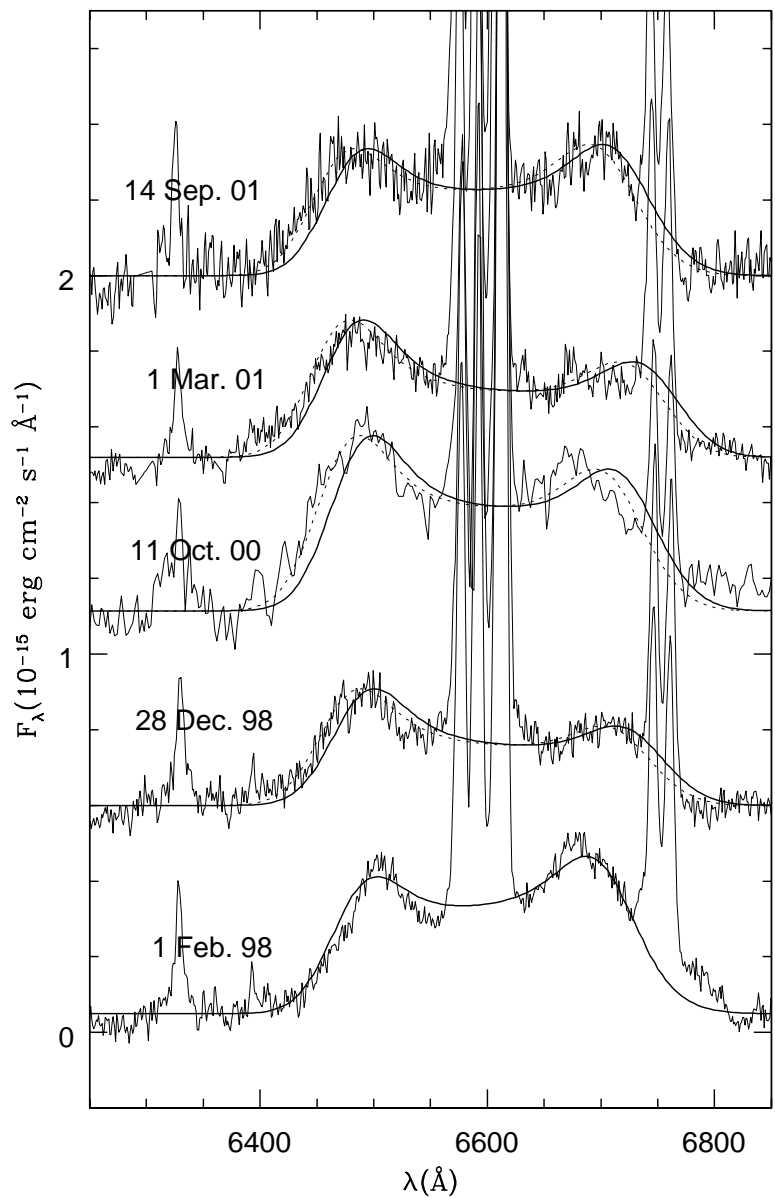}
\caption{Fits of the elliptical disk model to the profiles from
  1998 Feb. to 2001 Sep. by changing only $\phi_0$ and $\xi_q$.
  Dotted lines show these fits blueshifted by 12~\AA . \label{fig8}}
\end{figure}

To fit individual line profiles we fixed the inner and outer
pericenter distances of the line-emitting disk to $\xi_1=450$ and
$\xi_2=1600$ (respectively), the inclination angle to $i=34^\circ$, and
the broadening parameter to $\sigma=1200~{\rm km~s}^{-1}$. The
eccentricity was allowed to increase linearly with radius, from
$e(\xi_1)=0$ to $e(\xi_2)=0.45$. The emissivity of the disk
($\epsilon\propto \xi^{-q}$) was set to increase with radius in the
inner disk (i.e., $q_1=-1$ up to a radius $\xi_q$) and then decrease
($q_2=1$). The orientation angle of the disk, $\phi_0$, and the
emissivity break radius, $\xi_q$, were the only parameters allowed to
vary to represent a precession of the disk and a decline in the
illumination. We first searched for models that could reproduce the
primary observables, namely the velocities and relative strengths of
the two peaks. The best-fitting models resulting from the above
procedure are compared with the observations in Figures \ref{fig7} and
\ref{fig8}, and the model parameters are summarized in
Table~\ref{tbl-3}.

In the last epochs, beginning in 1998 Dec., the fits of the model to
the data show some improvement if we allow a small blueshift of the
central wavelength of the profile ($\sim$10\AA) relative to the wavelength of the
narrow H$\alpha$ line (6591.7~\AA\ --- which we have so far
adopted as the center of the broad line in the models).
We show these improved fits as dotted lines in Figure \ref{fig8}.
This blueshift of the broad line was already detected
in the observed evolution of the blue and red peak
velocities (Fig. \ref{fig4}),
and was discussed at the end of the previous subsection.

\begin{figure}
\plotone{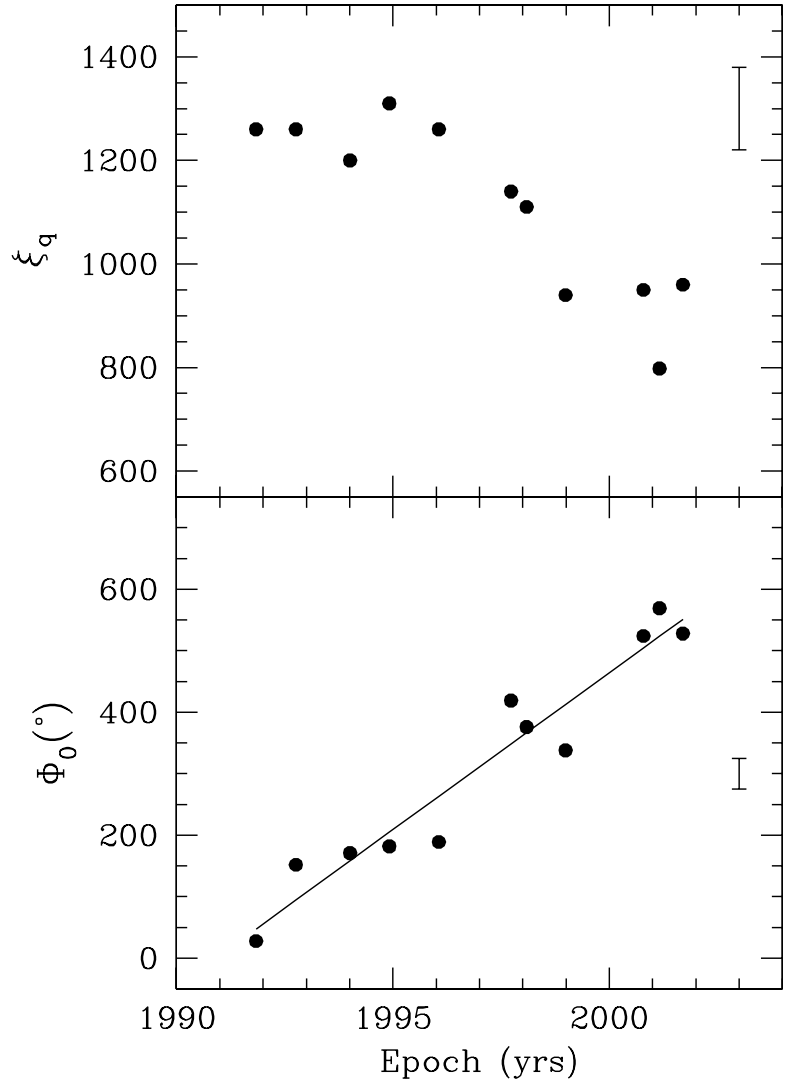}
\caption{Evolution of the elliptical disk model parameters:
  emissivity-law peak radius $\xi_q$ (top); and orientation of the
  disk relative to the line of sight $\phi_0$ (bottom).
\label{fig9}}
\end{figure}

\begin{figure}
\plottwo{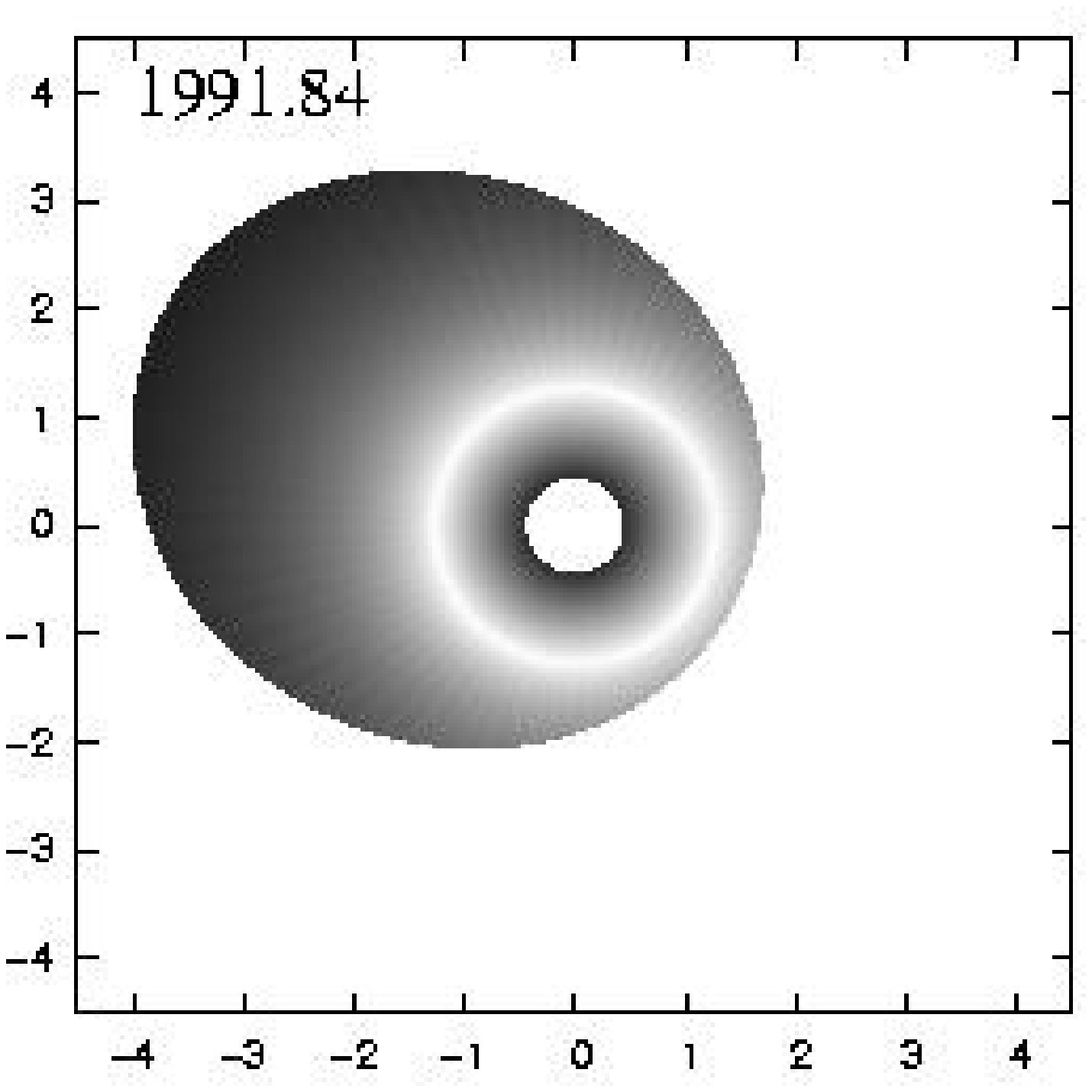}{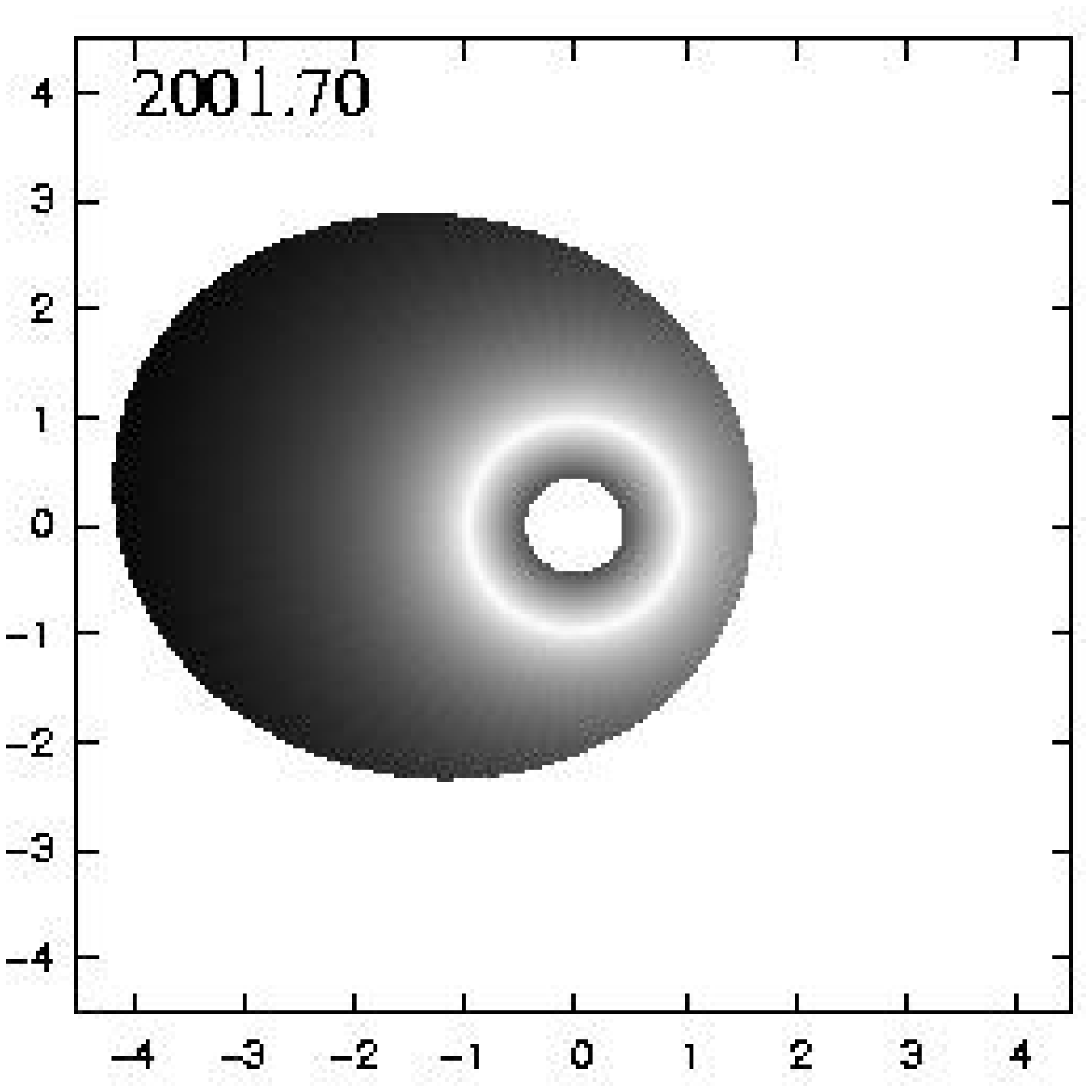}
\caption{Elliptical disk cartoon, illustrating the change in the disk emissivity and
  orientation between the first and last epoch fitted with this model.
  Units are 10$^3r_g$ and the observer is to the right. Emissivity scale is
  arbitrary, white representing the brightest regions.
\label{fig10}}
\end{figure}

The evolution of $\xi_q$ and $\phi_0$ are shown in
Figure \ref{fig9}. There is very little change in $\xi_q$ until 1996,
but then in 1997 it begins to decrease, causing the relative contribution
of the inner parts of the disk to become more important and
producing the broadening of the profile. The orientation angle of the disk,
represented by $\phi_0$, shows a small range of values,
$131 < \phi_0 < 209$, as listed in Table~\ref{tbl-3}. Noticing that
changes in orientation by half a revolution
result in very similar (although not identical) model profiles,
we have exploited this property in order to construct a series of models
that would approximately represent a precession of the disk.
The best compromise we could obtain between a good fit of the profiles and
the precession scenario is illustrated in the bottom panel of Figure \ref{fig9}.
Assuming that the precession rate of the disk is constant,
we carry out a linear least-squares fit to the precession
phase shown in Figure \ref{fig9}. We adopt a functional form given by
\begin{equation}
{\phi_0\over 2\pi} = {t-t_0\over P_{\rm prec}}\; ,
\label{eq_prec}
\end{equation}
where $t$ is the time in years, $t_0$ is an arbitrary zero point of
the precession cycle, and $P_{\rm prec}$ is the precession period.
The best fit, shown as a solid line in Figure \ref{fig9}, yields
$t_0=1990.8$ and $P_{\rm prec}=7.1$~years, which implies that the
disk has completed almost two revolutions since the epoch of the first
observation as one would expect from the behavior of the peak strength
ratio, $F_B/F_R$, shown in Figure \ref{fig5}. In Figure \ref{fig10} we show
a cartoon of the accretion disk in two epochs, the first (1991.8) and
last (2001.7)  fitted by the model, illustrating the change in the
disk emissivity in 10 years.

In conclusion, we note that even though models that represent a smooth
precession of the eccentric disk are preferable on physical grounds,
they do not provide as good a description of the line profiles as the
models whose parameters are listed in Table~\ref{tbl-3}.

\subsection{Circular Disk Plus Spiral Arm Model\label{s_sp}}

An alternative to the elliptical disk model is
a model that invokes a non-axisymmetric perturbation of the emissivity
of a circular disk. Models of this type were applied to the variable
H$\alpha$ profile of 3C~390.3 by Zheng, Veilleux, \& Grandi (1991; a
``bright sector'' in the disk) and by Gilbert et al. (1999; a
single-arm spiral, inspired by the work of Chakrabarti \& Wiita 1993,
1994). An important feature of these models is that the perturbation
spans a relatively wide range in radius (and projected velocity),
which leads to a modulation in the relative strengths of the
two peaks of the line profile without creating a ``third peak'' that
moves across the profile. Spiral arm models have additional appeal
because spiral arms are known to exist in other astrophysical disks,
such as the disks of spiral galaxies and the accretion disks of some
cataclysmic variables (e.g., Steeghs, Harlaftis, \& Horne 1997;
Baptista \& Catal\'an 2000; Patterson, Halpern, \& Shambrook 1993).

We adopt the formulation of Gilbert et al. (1999), according to which
the line-emitting portion of the disk is circular, has an inclination
angle $i$, and is bounded by inner and outer radii $\xi_1$ and $\xi_2$.
Superposed on the axisymmetric emissivity pattern ($\epsilon\propto
\xi^{-q}$) there is an emissivity perturbation which has the form of a
single spiral arm. We adopted a single-arm pattern after experimenting
with patterns having two or more spiral arms. We found that two or
more spiral arms could not reproduce the changing asymmetry of the
H$\alpha$ line profile of NGC~1097, a conclusion also reached by
Gilbert et al. (1999) for 3C~390.3.

The prescription for the total emissivity of the disk is given by
\begin{equation}
\epsilon (\xi,\phi) = \epsilon_0 \; \xi^{-q} \;
\left\{ 1 +
\frac{A}{2} \exp\left[-\frac{4\ln 2}{\delta^2}(\phi-\psi_0)^2\right]
+\frac{A}{2}\exp\left[-\frac{4\ln 2}{\delta^2} (2\pi-\phi+\psi_0)^2\right]
\right\} \; ,
\label{eq_spiral}
\end{equation}
where the two exponentials represent the decay of the emissivity of
the spiral arm with azimuthal distance on either side of the
ridge line. The parameter $\delta$ represents the azimuthal width
of the spiral arm (FWHM), while the quantity $\phi-\psi_0$ is the
azimuthal distance from the ridge of the spiral arm, which is defined
by
\begin{equation}
\psi_0=\phi_0 + {\log{(\xi / \xi_{sp})}\over{\tan p}} \; ,
\label{eq_ridge}
\end{equation}
where $p$ is the pitch angle and $\xi_{\rm sp}$ is the innermost
radius of the spiral arm (also used as a fiducial radius for
normalization; the spiral arm is set to begin at the outer radius of
the disk, $\xi_2$). The parameter $A$ represents the brightness
contrast between the spiral arm and the underlying, axisymmetric disk,
while $\phi_0$ sets the azimuthal orientation of the spiral pattern.

To fit individual line profiles, we followed the procedure described
in \S\ref{s_ell}.  We found that we could keep the inner and outer
radii of the disk at the values of the inner and outer pericenter
distances of the elliptical disk model, namely $\xi_1=450$ and
$\xi_2=1600$. We also kept the inclination at $i=34\arcdeg$ and the
broadening parameter at $\sigma=1200~{\rm km~s}^{-1}$. The only
parameters we needed to vary in order to reproduce the evolution of
the profiles were $q$, $A$, and $\phi_0$. The remaining spiral arm
parameters were kept at the following values: $\xi_{sp}=\xi_1$,
$p=-50^\circ$, and $\delta=70^\circ$.  As in the case of the elliptical
disk model, the variation of $\phi_0$ can explain the
modulation of the relative strengths of the two peaks of the H$\alpha$
line but not the increasing separation between them. Therefore, in the
spirit of the discussion of \S\ref{interp}, we allowed the emissivity
power-law index to vary so as to shift the region of maximum emission
to smaller radii.

\begin{figure}
\plotone{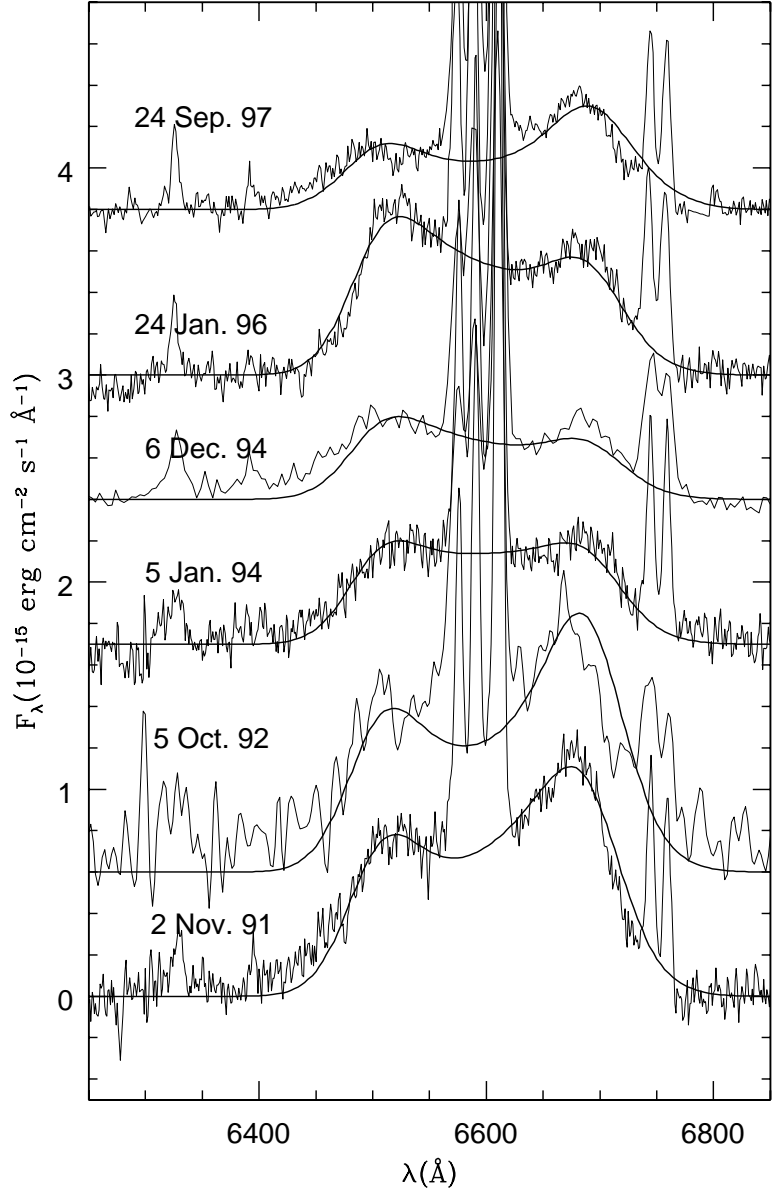}
\caption{Fits of the spiral disk model to the profiles from 1991 Nov.
  to 1997 Sep. by changing $\phi_0$ (the orientation of the spiral arm
  relative to the line of sight), $q$ (the slope of the radial
  component of the emissivity law), and $A$ (the spiral arm contrast
paramenter). The values of these parameters
  are listed in Table \ref{tbl-4}. \label{fig11}}
\end{figure}

\begin{figure}
\plotone{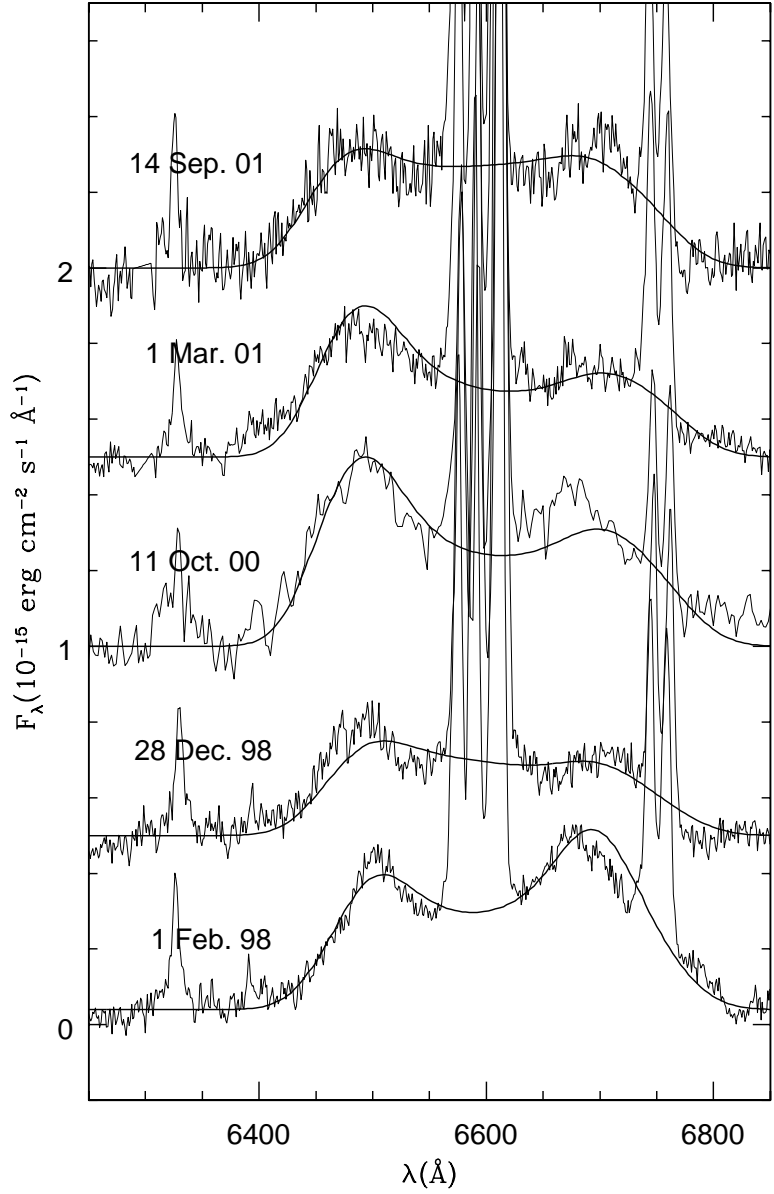}
\caption{Fits of the spiral disk model to the profiles from
  1998 Feb. to 2001 Sep. by changing $\phi_0$ and $q$. The values
  of these parameters are listed in Table~\ref{tbl-4}.\label{fig12}}
\end{figure}

\begin{figure}
\plotone{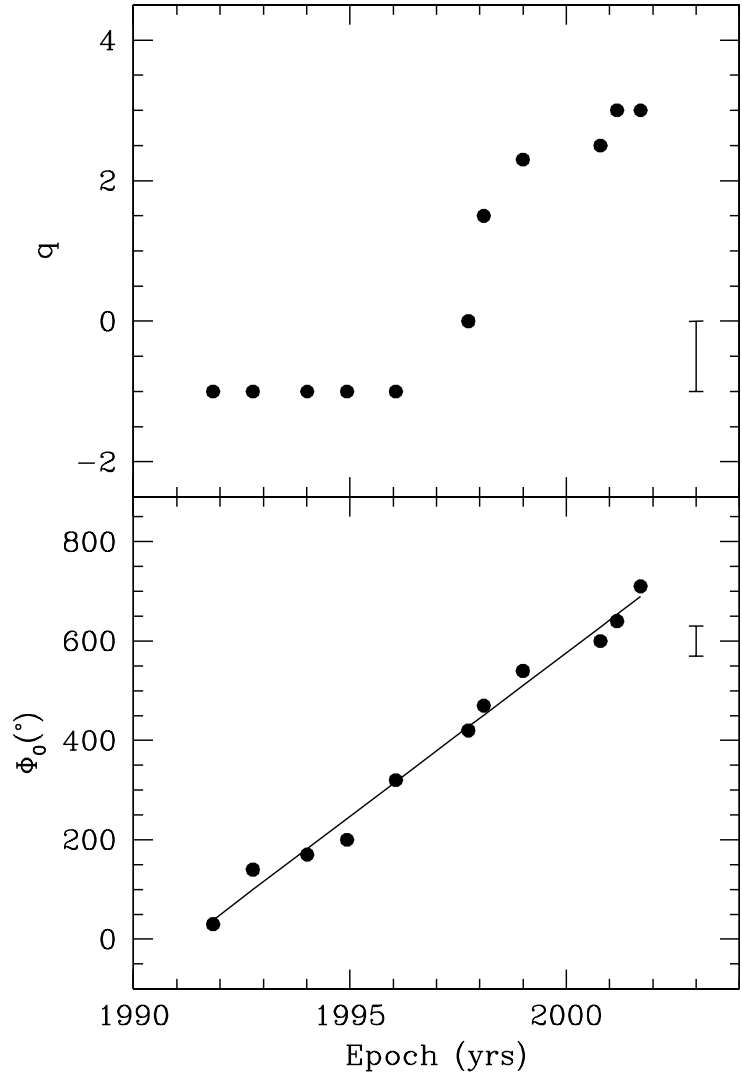}
\caption{Evolution of the spiral disk model parameters:
  emissivity-law power-law index $q$ (top), and orientation of
  the spiral arm relative to the line of sight $\phi_0$ (bottom)
  in the precessing scenario. The straight  line is a linear fit to
  the variation of $\phi_0$ with time.\label{fig13}}
\end{figure}

The best-fitting models are compared to the observed line profiles in
Figures \ref{fig11} and \ref{fig12}, while the corresponding parameter
values (for $A$, $q$, and $\phi_0$) are listed in Table \ref{tbl-4}.
The spiral arm contrast, parameterized by $A$, has
decreased from 3 to 2 since the first observations, while $q$ has increased,
favoring the emission of the inner part of the disk relative to that of the
outer parts in the later epochs and producing the broadening of the profile.
The $\phi_0$ values were constrained to allow
a monotonic precession of the spiral pattern, as illustrated in
Figure \ref{fig13}, where we show
how $q$ and $\phi_0$ evolve with time. The latter figure
shows that the spiral model accomodates better a
monotonic precession (of the spiral pattern)
than the elliptical model.
The solid line in the lower panel of Figure \ref{fig13} is a fit to
the precession phase following Eq. (\ref{eq_prec}),
yielding $t_0=1991.3$ and $P_{\rm prec}=5.5\pm 0.3$~years.
Such a period implies, again, that the
disk has completed almost two revolutions since the epoch of the first
observation as one would expect from the behavior of the peak strength
ratio ($F_B/F_R$) shown in Figure \ref{fig5}.

\begin{figure}
\vspace{12cm}
\plottwo{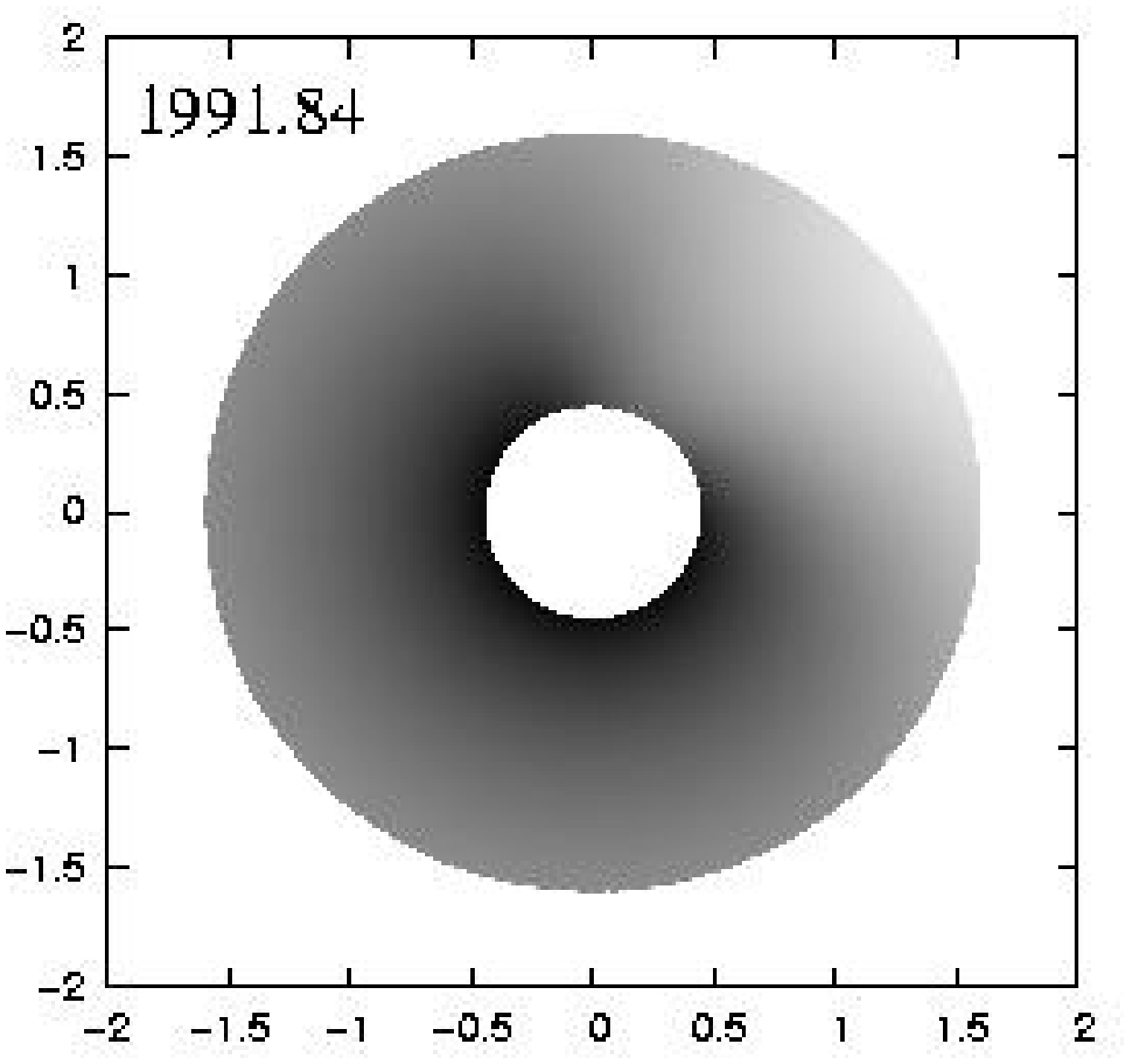}{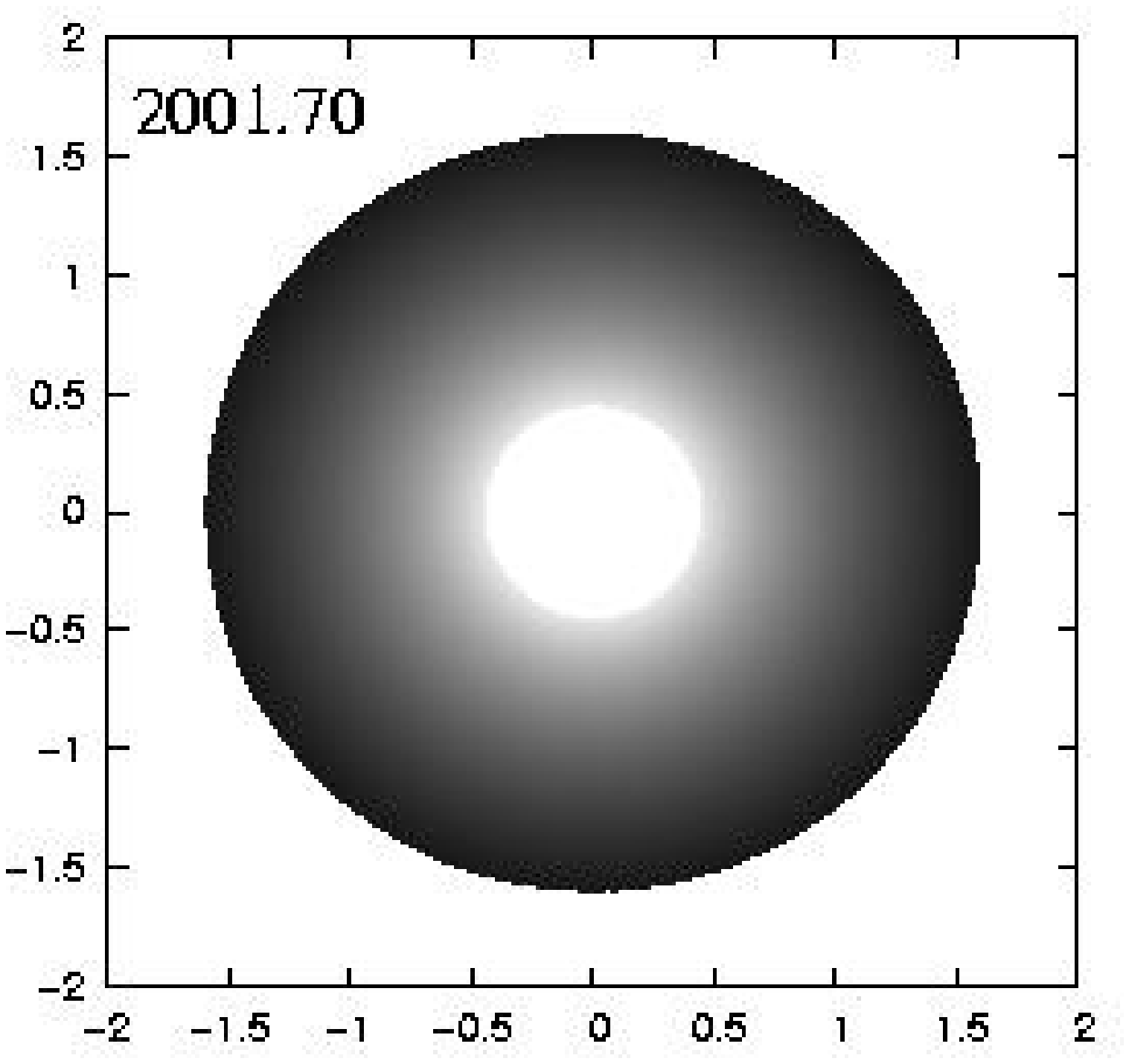}
\caption{Spiral disk cartoon, illustrating the disk emissivity and
  orientation for the first and last epoch fitted with this model. Units are
  10$^3r_g$ and the observer is to the right. Emissivity scale is arbitrary,
  white representing the brightest regions. \label{fig14}}
\end{figure}

In Figure \ref{fig14} we illustrate a cartoon of the disk in the first and last epochs
fitted by the model. Notice that the spiral arm is quite ``broad'' and
hard to observe in the last epoch due to the low contrast relative to the disk as
implied by the best-fitting parameters (Table \ref{tbl-4}).

\section{Discussion}

\subsection{Comparison With Studies of Other Objects}

The variation of the relative strengths of the two peaks of the
H$\alpha$ line of NGC~1097 appears to follow the same trend as in
broad-line radio galaxies. Examples include 3C~390.3 (Zheng, Veilleux,
\& Grandi 1991; Gilbert et al. 1999), 3C~332 (Gilbert et al. 1999),
Pictor~A (Eracleous \& Halpern 1998), and Arp~102B (Newman et al. 1997).
On the other hand, drifts in the peak wavelengths of double-peaked
profiles are seldom observed. One case previously reported is that of
the radio galaxy 3C~390.3, for which Gaskell (1996), Eracleous et
al. (1997), and Gilbert et al. (1999) found a drift in the velocity of
the blue peak between 1970 and 1996. But in 3C~390.3, Eracleous et
al. (1997) found that the two peaks moved together, unlike what
is observed in the case of NGC~1097, in which the two peaks move apart
from each other.

On the other hand, Shapovalova et al. (2001), using observations of
the H$\beta$ profile of 3C~390.3 from 1995 to 2000 (later than those
reported by Eracleous et al. 1997), claim to have found drifts of the
two peaks in opposite directions (or velocities),
arguing that the difference between
the velocities of the blue and red peaks is inversely correlated with the
continuum flux. They note that when the H$\beta$ flux decreases, the
radial velocity difference between the red and blue peaks increase,
while when the H$\beta$ flux increases, the velocity difference
between the two peaks decreases. We point out, however, that the
measurement of the wavelength of the red peak of the H$\beta$ line
in 3C~390.3 is quite uncertain because this peak is severely
contaminated by the [\ion{O}{3}] $\lambda\lambda$ 4959, 5007 emission
lines.

The behavior of the 3C~390.3 broad H$\beta$ profile as described by
Shapovalova et al. (2001) would be similar to the case of NGC~1097
reported here (as discussed in Sec.\ref{interp}), and they have
proposed a similar interpretation: the
flux increase of H$\beta$ is due to the increase in the source
luminosity, which thus ionizes regions farther from the nucleus (lower
velocity); the flux decrease of H$\beta$ is due to a decrease in the
ionizing luminosity, which shifts the region of maximum line emission
closer to the center of the disk (higher velocity).

\subsection{Implications of the Observational Results and Models\label{s_mass}}

The pattern precession period inferred from the model fits can be connected
to the mass of the central black hole. Although eccentric disks and
spiral arms are related structures (see, for example, the illustration
in Adams, Ruden, \& Shu 1989), in our application the precession rates
are set by different mechanisms.

In the case of an eccentric disk, the precession rate is controlled by
the relativistic advance of the pericenter.  Therefore, we adapt the
formula given by Weinberg (1972, p. 197), which we write as
\begin{equation}
P_{\rm prec}
= {2\pi\over 3}\; {1+e\over (1-e)^{3/2}}\; {GM\over c^3}\; {\xi}^{5/2}
= 10.4\; {1+e\over (1-e)^{3/2}}\; M_6\; {\xi_3}^{5/2}~{\rm years,}
\label{eq_adv}
\end{equation}
where $M_6$ is the mass of the black hole in units of $10^6\,{\rm
M}_{\odot}$ and ${\xi_3}$ is the pericenter distance of the
orbit in units of $10^3\; r_{\rm g}$. If the outer,
most eccentric ring of the disk is precessing as prescribed by the
advance of pericenter, then $\xi_3=1.6$, $e=0.45$, and, for a
period of 7.1~years obtained in \S\ref{s_ell}, the resulting
black hole mass is $6\times 10^4\,{\rm M}_{\odot}$. If
the inner rings impose a faster precession rate, the resulting
mass will be somewhat larger, with an upper limit corresponding
to the innermost radius of $\xi_3=0.45$ and $e=0$, which
yield a black hole mass of $5\times 10^6 \,{\rm M}_{\odot}$.

In the case of the precessing spiral pattern, we can connect the
precession period to the black hole mass using the results of Laughlin \&
Korchagin (1996), who find that the pattern period is several times
to an order of magnitude longer than the dynamical time. An extreme
upper bound to the pattern period is the sound-crossing time of the disk.
The dynamical and sound-crossing time scales are respectively given by
\begin{equation}
\tau_{\rm dyn} = 2\; M_6\; \xi_3^{3/2}~{\rm days}
\label{eq_tdyn}
\end{equation}
\begin{equation}
{\rm and} \quad \tau_{\rm s} = 8\; M_6\; \xi_3\; T_5^{-1/2}~{\rm months}
\label{eq_tsound}
\end{equation}
(adapted from Frank et al. 1992), where now $\xi_3$ is the radius in
the disk in units of $10^3\; r_{\rm g}$ and $T_5$ is the temperature
in units of $10^5$~K. By comparing these time scales with the
precession period of 5.5~years inferred in \S\ref{s_sp}, we estimate a
black hole mass in the range $10^7$--$10^8 \,{\rm M}_{\odot}$.

Another scenario discussed briefly in SB97 involves the precession of
a radiation-induced warp in the disk (Pringle 1996). In Eq. (4)
of SB97 we estimated a precession period of the order of a decade assuming a
black hole mass of $10^6 \,{\rm M}_{\odot}$ and a mass of the
accretion disk of $10^{-4} \,{\rm M}_{\odot}$. Since this disk mass
is, in fact, a lower limit (see SB97), the precession period
determined earlier implies an upper limit on the black hole mass of $M
< 10^6 \,{\rm M}_{\odot}$ in the context of this picture.

All of the above scenarios can be assessed using a black hole mass
derived from the velocity dispersion of the stars in the nucleus of
NGC~1097. Emsellem et al. (2001) measure a velocity dispersion of
$160\pm20~{\rm km~s}^{-1}$ in the inner 300~pc of the bulge of
NGC~1097.  Using the mass-velocity dispersion relation of Tremaine et
al. (2002), we infer a black hole mass in the range $(2-9) \times 10^7
\,{\rm M}_{\odot}$, where the uncertainty is dominated by the
dispersion about the best-fitting relation (see Figure 8 of Tremaine
et al.). This black hole mass disfavors both the elliptical disk and
precessing warp scenarios, leaving the spiral arm scenario as the best
candidate. The spiral arm scenario is also favored by the fact that it
provides the best description of the variable line profiles,
by better ``accommodating'' the precession scenario, as described
in \S4.

\section{Summary, Conclusions, and Open Questions}

We have used spectroscopic observations of the broad double-peaked
H$\alpha$ profile of the nucleus of NGC~1097 spanning the period
between 1991 Nov. and 2002 Oct. in order to constrain and
refine models for the variability of the line profile.  The profiles
display the following types of variations over these 11 years.

\begin{enumerate}

\item
A smooth, long-term variation in the relative strengths of the blue
and red peaks.

\item
A decrease in wavelength of the blue peak and a corresponding increase
in wavelength of the red peak --- in other words, a broadening of the
double-peaked profile, suggesting that the bulk of the emission is
shifting inward to gas that is moving at higher velocities.

\item
A decrease in the integrated luminosity of the line, so that in 2001
Sep. it was approximately 1/3 of its value of 1991 Nov.

\item
An inverse correlation between the width and luminosity of the line.

\item
A shift of the central wavelength of the profile (defined as the
average wavelength of the two peaks) from approximately zero velocity
relative to the narrow lines in 1991 to about $-450~{\rm
km~s}^{-1}$ in 2001.

\end{enumerate}

The evolution of the double-peaked H$\alpha$ profile of NGC~1097,
described above, lends additional support to our earlier suggestion
that the variations are a result of a precessing, non-axisymmetric
pattern in an accretion disk (Eracleous et al. 1995; SB95; SB97).
We have compared the observed line profiles with models of an eccentric
(elliptical) disk and a disk with a spiral arm and have shown that both models can
reproduce the data. However, we favor the spiral arm model
because (1) it can reproduce the evolution of the double-peaked
profile as due to the precession of the spiral arm, while the
elliptical disk model cannot reproduce a smooth precession; and (2) it
implies a black hole mass that is consistent with what is inferred from
the radial velocity dispersion of the stars in the nucleus, while the
elliptical disk model requires a mass which is an order of magnitude lower.

We have further found that the dimming of the broad double-peaked
H$\alpha$ line is coupled with the increasing separation of the blue
and red peaks. We have interpreted this behavior as a result of the
dimming of the source of ionizing radiation, which makes the
maximum emission region of the disk to drift inward,
and thus to regions of higher orbital velocities.
This dimming of the central source is progressing
independently of dynamical phenomena
in the disk, which is a reasonable outcome
of the tidal disruption scenario that explains
the abrupt appearance of the double-peaked lines.
The coupling between the luminosity and width of the line seems
to hold also at shorter time scales: an increase in the flux of the line
is associated with a decrease in the line width, while  a
decrease in flux is associated with an increase in line width.



In the process of studying the evolution of the H$\alpha$ profile and
fitting disk models to them we found a gradual blueshift, especially
at later epochs. This blueshift manifests itself as a gradual shift of
the average wavelength of the two peaks. We have not attempted to
model this effect since our data so far cannot yield strong
constraints. Nevertheless, we speculate that this blueshift may be
associated with the onset of an accretion disk wind (Elvis 2000).
Outflows with typical velocities of 1000\,km\,s$^{-1}$ --- consistent with the
blueshift of $\sim$400--500\,km\,s$^{-1}$ we have measured for NGC~1097
if we consider the inclination of $34^\circ$ for the disk --- have been
frequently observed in the spectra of Seyfert galaxies as absorption
lines in the UV (e.g., Kraemer, Crenshaw,
\& Gabel 2001) and X-rays (the ``warm absorber''), as well as in
the emission lines of the narrow-line region. Models for these
outflows attribute their origin in the accretion disk, as, for
example, a hydromagnetically driven wind of discrete clouds (Emmering,
Blandford, \& Shlosman 1992), or radiatively driven continuous winds
(Murray et al. 1995; Proga, Stone, \& Kallman 2000).  The presence of
an outflow in the case of NGC~1097 is also consistent with
observations of \ion{Fe}{2} absorption lines in a recent {\it HST} UV
spectrum (Eracleous 2002). Future observations of the variability
of the Balmer-line profiles should shed more light on the issue of
the possible wind.

\acknowledgements

We thank  Steinn Sigurdsson and Mario Livio for useful discussions on
issues related to accretion disk dynamics. We acknowledge
the Brazilian Institutions CNPq, CAPES, and FAPERGS for
partial support. This research was also supported in part by NASA
through grant NAG-81755 to the University of Maryland.
A.V.F.'s work is supported by NASA through grant GO-8684
from the Space Telescope Science
Institute, which is operated by the Association of Universities for
Research in Astronomy, Inc., under NASA contract NAS~5-26555.
We thank the staffs of the observatories used to obtain the
spectra in this paper, as well as many associates who helped with
the observations and reductions.
The W.~M. Keck Observatory is operated as a
scientific partnership among the California Institute of Technology,
the University of California, and NASA; the observatory was made
possible by the generous financial support of the W.~M. Keck
Foundation.


\clearpage

\begin{deluxetable}{llrcc}
\tablecaption{Observation Log\label{tbl-1}}
\tablewidth{0pt}
\tablehead{
\colhead{Date}       &                      & \colhead{Exposure} & \colhead{Aperture} & \colhead{Scaling} \\
\colhead{(mm/dd/yy)} & \colhead{Telescope}  & \colhead{(s)}      & \colhead{(\arcsec\ )} & \colhead{factor} \\
}
\startdata
  11/02/91       &  CTIO 4m       &   2700     & 2.0$\times$4.0 &1.12 \\
  10/05/92       &  CTIO 4m       &    685     & 2.0$\times$4.0 &1.32 \\
  09/12/93       &  Lick 3m       &   1200     & 2.0$\times$4.0 &0.52 \\
  01/05/94       &  CTIO 4m       &   1800     & 2.0$\times$4.0 &1.14 \\
  12/06/94       &  CTIO 4m       &   1200     & 1.5$\times$3.7 &1.46  \\
  01/24/96       &  CTIO 4m       &   1200     & 2.0$\times$4.0 &1.00 \\
  10/10/96       &  Lick 3m       &   3600     & 2.0$\times$4.7 &0.91 \\
  01/02/97       &  CTIO 1.5m     &   3600     & 1.8$\times$2.6 &0.66 \\
  09/24/97       &  CTIO 4m       &   1800     & 1.5$\times$3.7 &1.49 \\
  09/27/97       &  KPNO 2.1m     &   3600     & 1.9$\times$3.8 &1.22 \\
  01/02/98       &  CTIO 1.5m     &   6000     & 1.8$\times$2.6 &1.79 \\
  01/17/98       &  Keck II       &    500     & 1.0$\times$4.8 &1.28 \\
  02/01/98       &  CTIO 4m       &   1800     & 1.5$\times$3.7 &1.27 \\
  10/15/98       &  KPNO 2.1m     &   5400     & 1.9$\times$4.6 &1.42 \\
  10/20/98       &  CTIO 1.5m     &   5400     & 1.8$\times$6.5 &0.87 \\
  12/16/98       &  MDM 2.4m      &   2400     & 1.5$\times$2.9 &1.19 \\
  12/28/98       &  CTIO 4m       &   1800     & 1.5$\times$3.7 &1.29 \\
  01/06/99       &  Keck II       &    300     & 1.0$\times$2.4 &1.11 \\
  12/02/99       &  KPNO 2.1m     &   3600     & 1.9$\times$3.0 &1.21 \\
  09/23/00       &  KPNO 2.1m     &   3600     & 1.9$\times$3.0 &1.10\\
  10/11/00       &  Keck II       &    300     & 1.0$\times$2.4 &0.70 \\
  03/01/01       &  CTIO 4m       &    900     & 1.5$\times$3.7 &1.22\\
  09/14/01       &  NTT 3.6m      &   1800     & 1.5$\times$3.7 &1.33\\
  10/10/02       &  KPNO 2.1m     &   3600     & 1.9$\times$3.0 &1.17\\
\enddata
\end{deluxetable}


\begin{deluxetable}{crrrrr}
\tablecaption{Measured Properties
\label{tbl-2}}
\tablewidth{0pt}
\tablehead{
\colhead{ Epoch } &
\colhead{$\lambda_B$(\AA) } &
\colhead{$\lambda_R$(\AA)} &
\colhead{ $F_B$\tablenotemark{a}}  &
\colhead{$F_R$\tablenotemark{a}} &
\colhead{$F_{broad}$\tablenotemark{b}}
}
\startdata
 1991.84     &6528.0$\pm~3.0$  &6670.4$\pm~1.9$ &0.754$\pm~0.020$  &1.172$\pm~0.020$ &227.0$\pm~9.6$\\
 1992.76     &6535.9$\pm~2.4$  &6664.6$\pm~2.7$ &0.801$\pm~0.030$  &1.122$\pm~0.003$ &204.3$\pm~13.1$\\
 1994.01     &6530.9$\pm~0.4$  &6671.1$\pm~2.2$ &0.497$\pm~0.010$  &0.564$\pm~0.036$ &120.3$\pm~6.1$\\
 1994.92     &6519.3$\pm~2.2$  &6676.8$\pm~2.2$ &0.429$\pm~0.024$  &0.368$\pm~0.032$ &101.0$\pm~14.0$\\
 1996.06     &6528.1$\pm~0.5$  &6672.3$\pm~3.3$ &0.854$\pm~0.002$  &0.628$\pm~0.020$ &165.7$\pm~3.1$\\
 1997.73     &6508.6$\pm~5.5$  &6675.7$\pm~2.7$ &0.304$\pm~0.017$  &0.545$\pm~0.016$ &102.1$\pm~3.8$\\
 1998.07     &6504.7$\pm~4.4$  &6679.3$\pm~1.0$ &0.398$\pm~0.005$  &0.430$\pm~0.006$ &102.3$\pm~11.1$\\
 1998.09     &6510.1$\pm~2.5$  &6679.4$\pm~0.7$ &0.335$\pm~0.017$  &0.461$\pm~0.007$ &104.8$\pm~1.8$\\
 1998.80     &6502.1$\pm~2.5$  &6686.5$\pm~6.5$ &0.334$\pm~0.015$  &0.292$\pm~0.012$ &76.6$\pm~6.0$\\
 1998.99     &6496.6$\pm~2.3$  &6690.5$\pm~5.8$ &0.283$\pm~0.009$  &0.225$\pm~0.002$ &70.1$\pm~2.1$\\
 1999.02     &6494.5$\pm~2.2$  &6687.2$\pm~7.5$ &0.302$\pm~0.010$  &0.227$\pm~0.006$ &71.8$\pm~12.0$\\
 2000.73     &6491.8$\pm~2.5$  &6673.5$\pm~4.0$ &0.502$\pm~0.020$  &0.380$\pm~0.015$ &102.3$\pm~11.0$\\
 2000.79     &6493.1$\pm~2.3$  &6674.4$\pm~3.2$ &0.441$\pm~0.021$  &0.298$\pm~0.013$  &100.0$\pm~7.5$\\
 2001.16     &6495.1$\pm~1.6$  &6687.9$\pm~2.8$ &0.338$\pm~0.001$  &0.254$\pm~0.005$  &80.2$\pm~6.8$\\
 2001.70     &6496.4$\pm~2.0$  &6691.4$\pm~1.1$ &0.326$\pm~0.007$  &0.338$\pm~0.012$  &84.4$\pm~8.5$\\
\enddata
\tablenotetext{a}{Peak flux density (10$^{-15}$ erg~cm$^{-2}$~s$^{-1}$~\AA$^{-1}$).}
\tablenotetext{b}{Integrated flux (10$^{-15}$ erg~cm$^{-2}$~s$^{-1}$).}
\end{deluxetable}


\clearpage

\begin{deluxetable}{crr}
\tablecaption{Elliptical Disk Model Parameters\label{tbl-3}}
\tablewidth{0pt}
\tablehead{
\colhead{ Epoch } & \colhead{$\xi_q$} & \colhead{$\phi_0$($^\circ$)}}
\startdata
 1991.84  & 1260    & 152  \\
 1992.76  & 1260    & 152  \\
 1994.01  & 1200    & 171  \\
 1994.92  & 1310    & 182  \\
 1996.06  & 1260    & 189  \\
 1997.73  & 1140    & 131  \\
 1998.09  & 1110    & 164  \\
 1998.99  &  940    & 202  \\
 2000.79  &  950    & 187  \\
 2001.16  &  798    & 209  \\
 2001.70  &  960    & 168  \\
\enddata
\end{deluxetable}


\begin{deluxetable}{crrr}
\tablecaption{Spiral Disk Model Parameters\label{tbl-4}}
\tablewidth{0pt}
\tablehead{
\colhead{ Epoch } &
\colhead{$A$} &
\colhead{$q$} &
\colhead{$\phi_0$($^\circ$)}
}
\startdata
 1991.84     & 3  & --1.0  &  30 \\
 1992.76     & 3  & --1.0  & 140 \\
 1994.01     & 2  & --1.0  & 170 \\
 1994.92     & 2  & --1.0  & 200 \\
 1996.06     & 2  & --1.0  & 320 \\
 1997.73     & 3  &   0.0  & 420 \\
 1998.09     & 2  &   1.5  & 470 \\
 1998.99     & 2  &   2.3  & 540 \\
 2000.79     & 2  &   2.7  & 600 \\
 2001.16     & 2  &   3.0  & 640 \\
 2001.70     & 2  &   3.0  & 710 \\
\enddata
\end{deluxetable}



\clearpage

\clearpage

\begin{thebibliography}{}

\bibitem[]{863} Adams, F. C., Ruden, S. P., \& Shu, F. H. 1989, \apj, 347, 959

\bibitem[]{868} Baptista, R., \& Catal\'an, M. S. 2000, ApJ, 539, L55

\bibitem[]{865} Barth, A. J., Ho, L. C., Filippenko, A. V., Rix, H.-W.,
  \& Sargent. W. L. W. 2000, ApJ, 546, 205



\bibitem[]{876}Bower, G.  A., Wilson, A. S., Heckman, T. M., \& Richstone,
  D. O. 1996, AJ, 111, 1901

\bibitem[]{879}Chakrabarti, S. K., \& Wiita, P. J. 1993, \aap, 271, 216

\bibitem[]{881}Chakrabarti, S. K., \& Wiita, P. J. 1994, \apj, 434, 518



\bibitem[]{888}Dumont, A. M., \& Collin-Souffrin, S. 1990, \aap, 229, 313

\bibitem[]{890} Elvis, M. 2000, \apj, 545, 63

\bibitem[]{892}Emmering, R. T., Blandford, R. D., \& Shlosman, I. 1992,
  ApJ, 385, 460

\bibitem[]{895} Emsellem, E., Greusard, D., Combes, F., Friedli, D.,
Leon, S., P\'econtal, E., \& Wozniak, H. 2001, A\&A 366, 52

\bibitem[]{898} Eracleous, M. 2002, in Mass Outflow in AGNs,
  ed. M. Crenshaw, et al. (San Francisco: Astron. Soc. Pacific), p. 131



\bibitem[]{905}Eracleous, M., \& Halpern, J. P. 1998, ApJ, 505, 577


\bibitem[]{909}Eracleous, M., Halpern, J. P., Gilbert, A. M., Newman,
  J. A., \& Filippenko, A. V. 1997, \apj, 490, 216

\bibitem[]{912}Eracleous, M., Livio, M., Halpern, J. P., \&
  Storchi-Bergmann, T. 1995, \apj, 438, 610


\bibitem[]{917}Ferrarese, L., \& Merritt, D. 2000, ApJ, 539, L9

\bibitem[]{919} Frank, J., King, A. R., \& Raine, D. J. 1992, in
  Accretion Power in Astrophysics (Cambridge: Cambridge Univ. Press)


\bibitem[]{926}Gaskell, C. M. 1996, ApJ, 464, L107

\bibitem[]{928}Gebhardt, K., et al. 2000, ApJ, 539, L13

\bibitem[]{930} Gilbert, A. M., Eracleous, M., Filippenko, A. V.,
   \& Halpern, J. P. 1999, in Structure and Kinematics of Quasar
  Broad-Line Regions, ed. C. M., Gaskell et al.
  (San Francisco: Astron. Soc. Pacific Conf. Ser. 175), p. 189


\bibitem[]{934}Ho, L. C., et al. 2000, ApJ, 541, 120


\bibitem[]{937}Kraemer, S. B., Crenshaw, D. M., \& Gabel, J. R. 2001,
  ApJ, 557, 30

\bibitem[]{940} Laughlin, G., \& Korchagin, V. 1996, ApJ, 460, 855

\bibitem[]{942}Murray, N., et al. 1995, ApJ, 451, 498


\bibitem[]{948}Newman, J. A., Eracleous, M., Filippenko, A. V., \&
Halpern, J. P.  1997, ApJ, 490, 216

\bibitem[]{945}Patterson, J., Halpern, J. P., \& Shambrook, A. 1993, ApJ,
  419, 803

\bibitem[]{951} Pringle, J. E. 1996, MNRAS, 281, 357

\bibitem[]{953}Proga, D., Stone, J., \& Kallman, T. R. 2000, ApJ, 543, 686

\bibitem[]{955}Rees, M. J. 1988, Nature, 333, 523


\bibitem[]{959}Shapovalova, A. I., et al. 2001, A\&A, 376, 775

\bibitem[]{961}Shields, J. C., et al. 2000, ApJ, 534, L27

\bibitem[]{963}Steeghs, D., Harlaftis, E., \& Horne, K. 1997, MNRAS, 290,
  L28

\bibitem[]{966}Storchi-Bergmann, T., Baldwin, J. A., \& Wilson,
  A. S. 1993, \apj, 410, L11 (SB93)

\bibitem[]{969}Storchi-Bergmann, T., Eracleous, M., Livio, M., Wilson,
  A. S., Filippenko, A. V., \& Halpern, J. P. 1995, \apj, 443, 617 (SB95)

\bibitem[]{972}Storchi-Bergmann, T., Eracleous, M., Ruiz, M. T., Livio,
  M., Wilson, A. S., \& Filippenko, A. V. 1997, \apj, 489, 87 (SB97)

\bibitem[]{975} Tremaine, S., et al. 2002, ApJ, 574, 740



\bibitem[]{982}Weinberg, S. 1972, Gravitation and Cosmology: Principles
  and Applications of the General Theory of Relativity (New York:
  J. Wiley \& Sons)




\bibitem[]{991}Zheng, W., Veilleux, S., \& Grandi, S. 1991, \apj, 381, 418



\end{thebibliography}
\end{document}